\documentclass[onecolumn,amsmath,showpacs,nofootinbib,12pt]{revtex4-1}
\usepackage{graphicx}
\usepackage{dcolumn}
\usepackage{bm}
\usepackage{color}
\usepackage[colorlinks=true,citecolor=darkred,urlcolor=darkred, pdfborder={0 0 0}]{hyperref}
\begin{document}
\newcommand{\hs}{\hspace*{0.5cm}}
\newcommand{\vs}{\vspace*{0.5cm}}
\newcommand{\be}{\begin{equation}}
\newcommand{\ee}{\end{equation}}
\newcommand{\bea}{\begin{eqnarray}}
\newcommand{\eea}{\end{eqnarray}}
\newcommand{\ben}{\begin{enumerate}}
\newcommand{\een}{\end{enumerate}}
\newcommand{\bde}{\begin{widetext}}
\newcommand{\ede}{\end{widetext}}
\newcommand{\nn}{\nonumber}
\newcommand{\crn}{\nonumber \\}
\newcommand{\Tr}{\mathrm{Tr}}
\newcommand{\non}{\nonumber}
\newcommand{\noi}{\noindent}
\newcommand{\al}{\alpha}
\newcommand{\la}{\lambda}
\newcommand{\bet}{\beta}
\newcommand{\ga}{\gamma}
\newcommand{\va}{\varphi}
\newcommand{\om}{\omega}
\newcommand{\pa}{\partial}
\newcommand{\+}{\dagger}
\newcommand{\fr}{\frac}
\newcommand{\bc}{\begin{center}}
\newcommand{\ec}{\end{center}}
\newcommand{\Ga}{\Gamma}
\newcommand{\de}{\delta}
\newcommand{\De}{\Delta}
\newcommand{\ep}{\epsilon}
\newcommand{\varep}{\varepsilon}
\newcommand{\ka}{\kappa}
\newcommand{\La}{\Lambda}
\newcommand{\si}{\sigma}
\newcommand{\Si}{\Sigma}
\newcommand{\ta}{\tau}
\newcommand{\up}{\upsilon}
\newcommand{\Up}{\Upsilon}
\newcommand{\ze}{\zeta}
\newcommand{\ps}{\psi}
\newcommand{\Ps}{\Psi}
\newcommand{\ph}{\phi}
\newcommand{\vph}{\varphi}
\newcommand{\Ph}{\Phi}
\newcommand{\Om}{\Omega}
\newcommand {\red} {\color{red}}
\newcommand {\blue} {\color{blue}}
\newcommand {\green} {\color{green}}
\definecolor{darkred}{rgb}{0.6,0,0}

\renewcommand\thesection{\arabic{section}}
\renewcommand\thesubsection{\arabic{section}.\arabic{subsection}}

\definecolor{linkcolor}{rgb}{0,0,0.5}
\renewcommand{\baselinestretch}{1.15}%
\title{The Dark Side of Flipped Trinification}    

\author{P. V. Dong}\email{pvdong@iop.vast.ac.vn}\affiliation{Institute of Physics, Vietnam Academy of Science and Technology, 10 Dao Tan, Ba Dinh, Hanoi, Vietnam}
\author{D. T. Huong}\email{dthuong@iop.vast.ac.vn}\affiliation{Institute of Physics, Vietnam Academy of Science and Technology, 10 Dao Tan, Ba Dinh, Hanoi, Vietnam}
\author{Farinaldo S. Queiroz}\email{queiroz@mpi-hd.mpg.de}\affiliation{Max-Planck-Institut f\"ur Kernphysik, Saupfercheckweg 1, 69117 Heidelberg, Germany\\
International Institute of Physics, Federal University of Rio Grande do Norte, Campus Universit\'ario, Lagoa Nova, Natal-RN 59078-970, Brazil
}
\author{Jos\'e W. F. Valle}\email{valle@ific.uv.es} \affiliation{AHEP Group, Instituto de F\'isica Corpuscular – C.S.I.C./Universitat de Valencia
Edificio de Institutos de Paterna, C/Catedratico Jos\'e
Beltran, 2 E-46980 Paterna (Valencia) - SPAIN}
\author{C. A. Vaquera-Araujo}\email{vaquera@fisica.ugto.mx} 
\affiliation{
Departamento de F\'isica, DCI, Campus Le\'on, Universidad de
Guanajuato, Loma del Bosque 103, Lomas del Campestre C.P. 37150, Le\'on, Guanajuato, M\'exico}
\affiliation{
Consejo Nacional de Ciencia y Tecnolog\'ia, Av. Insurgentes Sur 1582. Colonia Cr\'edito Constructor, Del. Benito Ju\'arez, C.P. 03940, Ciudad de M\'exico, M\'exico}

\begin{abstract}

  We propose a model which unifies the Left-Right symmetry with the
  $SU(3)_L$ gauge group, called flipped trinification, and based on
  the $SU(3)_C\otimes SU(3)_L\otimes SU(3)_R\otimes U(1)_X$ gauge
  group. The model inherits the interesting features of both
  symmetries while elegantly explaining the origin of the matter
  parity, $W_P=(-1)^{3(B-L)+2s}$, and dark matter stability. We
  develop the details of the spontaneous symmetry breaking mechanism
  in the model, determining the relevant mass eigenstates, and showing
  how neutrino masses are easily generated via the seesaw mechanism.
  Viable dark matter candidates can either be a fermion, a scalar
  or a vector, leading to potentially different dark matter
  phenomenology.
 
\end{abstract}

\date{\today}

\maketitle

\section{\label{introduction} Introduction}

The mystery of Dark Matter (DM) is one of the biggest open questions
in science
\cite{Bertone:2010,Queiroz:2016sxf,Capdevilla:2017doz,Kavanagh:2017hcl}.
Despite the fact that its existence has been ascertained at several
distance scales of our universe, its nature has not yet been resolved
and the Standard Model (SM) fails to account for it. The need to
extend the SM goes beyond the DM problem, due to the existence of
important open questions connected to neutrino masses, the
cosmological baryon-number asymmetry, inflation and
reheating. Besides, from the theoretical side, the SM fails to explain
the existence of (just) three fermion families as well as the origin
of the observed parity violation of the weak interaction. The purpose
of this paper is to study how an extension of the SM addressing these
two issues, while hosting a viable DM candidate.

The minimal left-right symmetric model based on the
$SU(3)_C \otimes SU(2)_L \otimes SU(2)_R \otimes U(1)_{B-L}$ gauge
group, completed by a $Z_2$ symmetry that interchanges the left and
right, is one of the most attractive extensions of the
SM~\cite{Pati:1974yy,Mohapatra:1974hk,Senjanovic:1975rk,minkowski:1977sc,Senjanovic:1978ev,Mohapatra:1979ia}.
It gives a manifest understanding for the origin of parity violation
in the weak interaction, neutrino mass generation as well as a
framework for dark
matter~\cite{Nemevsek:2012cd,Heeck:2015qra,Garcia-Cely:2015quu,Patra:2015vmp}.

By the same token, models based on the
$SU(3)_C \otimes SU(3)_L \otimes U(1)_N$ gauge group, for short 3-3-1,
offer plausible explanations for the number of generations and a
hospitable scenario for neutrino mass generation as well as
implementing a viable dark sector
\cite{Singer:1980sw,valle:1983dk,Pisano:1991ee,Frampton:1992wt,Foot:1992rh,Montero:1992jk,Foot:1992rh,Foot:1994ym,Hoang:1995vq,Reig:2016vtf}.
 
Hence it is theoretically well motivated to build a model where both
groups are described in a unified way. 
Indeed, models have been proposed in the context of the
$SU(3)_C \otimes SU(3)_L \otimes SU(3)_R$ gauge group
\cite{Babu:1985gi,Nishimura:1988fp,Carlson:1992ew,Willenbrock:2003ca,Sayre:2006ma,Cauet:2010ng,Dias:2010vt,Stech:2014tla,Pelaggi:2015kna,Ferreira:2015wja,Pelaggi:2015knk,Rodriguez:2016cgr,Camargo-Molina:2016bwm,Huong:2016kpa,Dong:2016sat,Reig:2016tuk,Hati:2017aez}. Since
they are based on a three copies of the $SU(3)$ non-Abelian group, it
has been coined the term trinification. 
The motivation for trinification lies in the unified description of
both strong and electroweak interactions using the same non-Abelian
gauge group, while incorporating nice features of both left-right and
3-3-1 gauge groups.  
Fully realistic models unifying left–right and 331 electroweak
symmetries have, in fact, been recently proposed using a flipped
trinification scenario with an extra $U(1)_X$
factor~\cite{Reig:2016tuk,Hati:2017aez}.

In this paper we focus on an interesting question, namely, can we
build a model preserving the nice features of the left-right and 3-3-1
symmetries while naturally explaining the origin of the matter parity
and dark matter?
We argue that, using the gauge principle to extend the trinification
framework, there is a compelling and minimal solution incorporating
dark matter and realistic fermion masses. Such a flipped trinification
setup is better motivated because inherits the good features of both
left-right and $SU(3)_L \otimes U(1)_N$ symmetries and, in addition,
elegantly addresses the origin of matter parity and dark matter
stability in the context of 3-3-1 type models
\cite{Mizukoshi:2010ky,Profumo:2013sca,Dong:2013wca,Dong:2013ioa,Kelso:2013nwa,Dong:2014esa,Dong:2015rka,Cogollo:2014jia,Dong:2014wsa,Kelso:2014qka,Alves:2015pea,Mambrini:2015sia,Huong:2016ybt},
while generating fermion masses with a minimal scalar sector.
Indeed, it suffices to have one triplet ($\chi_L$), one bitriplet
$(\phi)$, one sextet $(\sigma_R)$ to generate realistic fermion
masses, as opposed to earlier versions where another bitriplet was
necessary \cite{Dong:2016sat,Reig:2016tuk}.  In order to ensure
left-right symmetry further copies of the scalar multiplets are
required.
Thus, a minimal version of trinification with exact left-right
symmetry requires one bitriplet $(\phi)$, two sextets ($\sigma_L$ and
$\sigma_R$) and two triplets ($\chi_L$ and $\chi_R$).

The rest of this paper is organized as follows: In Sec. \ref{model},
we introduce the model with the gauge symmetry and particle content,
focusing on the particles with unusual $B-L$ charges. We find the
viable patterns of symmetry breaking and show that $W$-parity is a
residual gauge symmetry which protects the dark matter stability. In
Sec. \ref{mass}, we identify the physical fields and the corresponding
masses. In Sec. \ref{darkmater}, we present detailed calculations of
the dark matter observables. Finally, we summarize the results and
conclude this work in Sec. \ref{conclusion}.

\section{\label{model} A flipped trinification setup}

\subsection{Gauge Symmetry}

Trinification is a theory of unified interactions based on the gauge
symmetry $SU(3)_C\otimes SU(3)_L\otimes SU(3)_R$, the maximal subgroup
of $E_6$ \cite{Babu:1985gi,Nishimura:1988fp,Carlson:1992ew}. When
multiplied by an Abelian group factor, $U(1)_X$, we have the flipped
trinification~\cite{Reig:2016tuk,Hati:2017aez}, \be SU(3)_C\otimes
SU(3)_L\otimes SU(3)_R\otimes U(1)_X.\ee This symmetry can be obtained
by left-right symmetrizing the 3-3-1 model in order to account for
weak parity violation and close both $B-L$ and 3-3-1 algebras
(cf. \cite{Dong:2015yra}). An alternative motivation is that it can be
achieved from the minimal left-right symmetric model by enlarging the
left and right weak isospin groups in order to resolve the number of
fermion generations and accommodate dark matter
(cf. \cite{Dong:2016sat}).

The electric charge operator is generally given by 
\be
Q = T_{3L}+T_{3R}+ \beta (T_{8L}+T_{8R})+X,
\ee
which reflects the left-right symmetry, where $T_{nL,R}$
$(n=1,2,3,...,8)$ and $X$ are the $SU(3)_{L,R}$ and $U(1)_X$
generators, respectively. 
Note that $\beta$ is an arbitrary coefficient whose values dictate the
electric charge of the new fermions present in the model.

As usual, the baryon minus lepton number is embedded as
$Q=T_{3L}+T_{3R}+\fr 1 2 (B-L)$, which implies that 
\be B-L=2[\beta
(T_{8L}+T_{8R})+X]\ee 
is a residual gauge symmetry of
$SU(3)_L\otimes SU(3)_R\otimes U(1)_X$. Let us note that $B-L$ and
$SU(3)_L$ neither commute nor close algebraically. Therefore, the
present framework, along the 3-3-1-1 gauge theory, constitute a class
of models with a fully consistent formulation of gauged $B-L$ symmetry
in 3-3-1 extensions of the Standard Model
\cite{Dong:2013wca,Dong:2014wsa,Huong:2015dwa,Huong:2016ybt,Dong:2015yra}.

\subsection{Fermion Sector}
  
The fermion content in this model results simply from the left-right
symmetrization the left-handed fermion sector of the 3-3-1 model, so
as to produce the right-handed fermion sector. The fermion sector is
given as
  \be \psi_{aL} =
\left(\begin{array}{c}
               \nu_{aL}\\ e_{aL}\\ N^q_{aL}
\end{array}\right) \sim \left(1,3, 1,\frac{q-1}{3}\right),  \hs \hs  \psi_{aR} =
\left(\begin{array}{c}
               \nu_{aR}\\ e_{aR}\\ N^q_{aR}
\end{array}\right) \sim \left(1,1, 3,\frac{q-1}{3}\right),\ee
\be Q_{\al L}=\left(\begin{array}{c}
  d_{\al L}\\  -u_{\al L}\\  J^{-q-\frac{1}{3}}_{\al L}
\end{array}\right)\sim \left(3,3^*,1,-\frac{q}{3}\right), \hs \hs
Q_{\al R}=\left(\begin{array}{c}
  d_{\al R}\\  -u_{\al R}\\  J^{-q-\frac{1}{3}}_{\al R}
\end{array}\right)\sim \left(3,1,3^*,-\frac{q}{3}\right), \ee 
  \be  Q_{3L}= \left(\begin{array}{c} u_{3L}\\  d_{3L}\\ J^{q+\frac{2}{3}}_{3L} \end{array}\right)\sim
 \left(3,3,1,\frac{q+1}{3}\right), \hs \hs  Q_{3R}= \left(\begin{array}{c} u_{3R}\\  d_{3R}\\ J^{q+\frac{2}{3}}_{3R} \end{array}\right)\sim
 \left(3,1,3,\frac{q+1}{3}\right),\ee
 where $a=1,2,3$ and $\al=1,2$ are generation indices, and
 $q\equiv -(1+\sqrt{3}\beta)/2$.

 The new fields $N_a$ and $J_a$ above are new leptons and quarks
 predicted by the model. It can be easily shown that all triangle
 anomalies vanish, since both $SU(3)_L$ or $SU(3)_R$ groups match the
 number of fermion generations to be that of fundamental colors,
 in agreement with the current observations
 \cite{Patrignani:2016xqp}. This choice of fermion representations is
 the minimal for a flipped trinification
 \cite{Babu:1985gi,Nishimura:1988fp,Carlson:1992ew}.

\subsection{Scalar Sector}

To break the gauge symmetry and generate the masses properly, we need introduce the scalar multiplets as follows, 
\bea
\phi &=& \left(
\begin{array}{ccc}
 \phi_{ 11}^0 & \phi_{ 12}^+& \phi_{ 13}^{-q} \\
  \phi_{21}^- & \phi_{ 22}^{0} & \phi_{ 23}^{-1-q} \\
  \phi_{ 31}^{q}& \phi_{ 32}^{1+q}& \phi_{ 33}^{0} \\
\end{array}
\right)     \sim (1,3,3^*,0),\\
 \chi_L &=& \left( \begin{array}{ccc}\chi_1^{-q}\\ \chi_2^{-q-1}\\ \chi_3^0 \end{array} \right)_L \sim\left(1,3,1,-\frac{2q+1}{3} \right), \\
 \chi_R &=& \left( \begin{array}{ccc}\chi_1^{-q}\\ \chi_2^{-q-1}\\ \chi_3^0 \end{array} \right)_R \sim\left(1,1,3,-\frac{2q+1}{3} \right), \\
 \sigma_L &=& \left(%
\begin{array}{ccc}
 \sigma_{11}^0 & \frac{\sigma_{12} ^-}{\sqrt{2}}& \frac{\sigma_{ 13}^{q}}{\sqrt{2}} \\
  \frac{\sigma_{12}^-}{\sqrt{2}} & \sigma_{22}^{--} &\frac{ \sigma_{23}^{q-1}}{\sqrt{2}} \\
 \frac{ \sigma_{13}^{q}}{\sqrt{2}}&\frac{ \sigma_{23}^{q-1}}{\sqrt{2}}& \sigma_{33}^{2q} \\
\end{array}%
\right)_L     \sim \left(1,6,1,\frac{2(q-1)}{3}\right), \\
\sigma_R &=& \left(%
\begin{array}{ccc}
 \sigma_{11}^0 & \frac{\sigma_{12} ^-}{\sqrt{2}}& \frac{\sigma_{ 13}^{q}}{\sqrt{2}} \\
  \frac{\sigma_{12}^-}{\sqrt{2}} & \sigma_{22}^{--} &\frac{ \sigma_{23}^{q-1}}{\sqrt{2}} \\
 \frac{ \sigma_{13}^{q}}{\sqrt{2}}&\frac{ \sigma_{23}^{q-1}}{\sqrt{2}}& \sigma_{33}^{2q} \\
\end{array}%
\right)_R     \sim \left(1,1,6,\frac{2(q-1)}{3}\right),\eea
with the corresponding VEVs, 
\bea
\langle\phi \rangle&=&\frac{1}{\sqrt{2}} \left(%
\begin{array}{ccc}
u & 0 &0\\
 0 & u^\prime &0 \\
0& 0&w \\
\end{array}%
\right), \hs  \langle \chi_R \rangle =\frac{1}{\sqrt{2}} \left(%
\begin{array}{ccc}
0\\
 0 \\
w^\prime \\
\end{array}%
\right), \hs \langle \si_R \rangle =\frac{1}{\sqrt{2}} \left(%
\begin{array}{ccc}
\Lambda & 0 &0\\
 0 & 0&0 \\
0& 0&0 \\
\end{array}%
\right).
\label{vev}\eea

Note that the scalars transform as $\phi \to U_L\phi U^\dagger_R$,
$\chi_R \to U_R\chi_R $, and $\sigma_R \to U_R \sigma_R U^T_R$ under
$SU(3)_L\otimes SU(3)_R$. We emphasize that these three scalar
multiplets are sufficient to generate all fermion masses. The scalar
multiplets $\chi_L$ and $\sigma_L$ have been added to ensure the
left-right symmetry, but they do not play any role in our
phenomenology because the VEV of these fields are neglible hence
contributing neither to gauge boson masses nor to the spontaneous
symmetry breaking pattern~\footnote{They only contribute to the tiny
  neutrino masses.}.
Therefore, for simplicity hereafter we ignore the VEVs of
$\sigma_L,\chi_L$, keeping only the VEVs of $\sigma_R,\chi_R$, denoted
omitting the subscript ``$R$''. We now discuss what types of
spontaneous symmetry breaking patterns one may have in our model.

\subsection{Spontaneous Symmetry Breaking}

We now address the issue of which types of symmetry breaking patterns
can be achieved within our model.

\subsubsection{Case 1: $w,w'\gg \La \gg u, u^\prime$}

In this scenario, we assume $w,w'\gg \La \gg u, u^\prime$, leading to
the following breaking pattern, 
\bc
\begin{tabular}{c}  
$SU(3)_C\otimes SU(3)_L\otimes SU(3)_R \otimes U(1)_X$\\
$  \downarrow w, w^\prime$\\
$SU(3)_C\otimes SU(2)_L \otimes SU(2)_R \otimes U(1)_{B-L}$\\
$\downarrow\Lambda$ \\
$SU(3)_C\otimes SU(2)_L \otimes U(1)_Y \otimes W_P$\\
$\downarrow u, u^\prime$\\ 
$SU(3)_C\otimes U(1)_Q\otimes W_P.$
\end{tabular}
\ec  

Notice that spontaneous symmetry breaking leaves the residual discrete
gauge symmetry, $W_P$, conserved along with the electric and color
charges. Let us now identify what symmetry is that.
The VEV of $\sigma_{11}^0$, $\La$, breaks $B-L$ since
$[B-L]\langle \sigma^0_{11}\rangle=\sqrt{2}\La\neq 0$, where
$\sigma^0_{11}$ has $B-L=2$. The $U(1)_{B-L}$ transformation that
preserves the vacuum is
$\langle \sigma^0_{11}\rangle \rightarrow e^{i\omega(B-L)}\langle
\sigma^0_{11}\rangle =e^{i2\omega}\langle \sigma^0_{11}\rangle=\langle
\sigma^0_{11}\rangle$, with $\omega$ as a transformation parameter.

Thus, we obtain $e^{i2\omega}=1$, or $\omega = m \pi$ for
$m= 0, \pm 1, \pm 2,...$, and the surviving transformation is
$M_P=e^{im\pi (B-L)}=(-1)^{m(B-L)}$. Since the spin parity $(-1)^{2s}$
is always conserved due to Lorentz symmetry, the residual discrete
symmetry preserved after spontaneous symmetry breaking is
$W_P=M_P\times (-1)^{2s}$, which is actually a whole class of
symmetries parameterized by $m$. Among such conserving
transformations, we focus on the one with $m=3$,
\be
W_P=(-1)^{3(B-L)+2s},\ee 
which we call the matter parity~\footnote{ We note that the matter
  parity present in our model coincides with $R$-parity in
  supersymmetry.}. We stress that in our model, it emerges as a
residual gauge symmetry,
\be W_P=(-1)^{6[\beta(T_{8L}+T_{8R})+X]+2s}, \ee 
and it acts nontrivially on the fields with unusual (wrong) $B-L$
numbers. For details, see Table \ref{tb1}. $W$-parity, $W_P$, is thus
named following the ``wrong'' item as in previous studies.

\subsubsection{Case 2: $\La \gg w,w^\prime \gg u,u^\prime$}

For $\La\gg w,w'$, the gauge symmetry is broken following a different path, 
\begin{eqnarray*}
    &
    SU(3)_{C}^{}\otimes SU(3)_{L}^{}\otimes SU(3)_{R}^{}\otimes U(1)_{X}^{} &\nonumber\\
    &
    \hspace*{-0.8cm} \downarrow \hspace*{-0.0cm} \Lambda  &\nonumber\\
    &
    SU(3)_{C}^{}\otimes SU(3)_{L}^{}\otimes SU(2)_{R'}^{}\otimes U(1)_{X'}^{}\otimes W'_{P} \nonumber\\
    &
    \hspace*{-1.0cm} \downarrow \hspace*{-0.0cm} w, w^\prime  &\nonumber\\
    &
    SU(3)_{C}^{}\otimes SU(2)_{L}^{}\otimes U(1)_{Y}\otimes W_P \nonumber\\
    &
    \hspace*{0.1cm}\downarrow   u,u'   &\nonumber\\
    &\hspace*{-1.0cm} SU(3)_C^{}  \otimes U(1)_{Q}\otimes W_P\,.&
\end{eqnarray*}

The $SU(2)_{R'}$ symmetry is generated by
$\{T_{6R},\ T_{7R},\ \frac{1}{2}(\sqrt{3}T_{8R}-T_{3R})\}$, meaning
that the left-right symmetry is initially broken in this case. The
$U(1)_{X'}$ charge is
$X'=\frac{\sqrt{3}+\beta}{4}(T_{8R}+\sqrt{3}T_{3R})+X$, with
$\beta=-(1+2q)/\sqrt{3}$. The discrete symmetry $W'_{P}$ takes a form,
$W'_{P}=(-1)^{m(\beta T_{8R}+X)} $, which is a residual symmetry of a
broken $U(1)$ group, with $U(\omega)=e^{i\omega 2(\beta T_{8R}+X)}$
transformation. The second stage of the symmetry breaking is driven
by $\ph_{33}^0, \chi^0_3$ fields. The VEV of $\chi^0_3$ breaks the
symmetry $SU(2)_{R'} \otimes U(1)_{X^\prime}$, while the VEV of
$\phi^0_{33}$ breaks not only that symmetry but also $W'_{P}$ and a
$U(1)$ group, with $U(\omega')=e^{i\omega' 2\beta T_{8 L}}$
transformation, as a $SU(3)_L$ subgroup. However, the VEV of
$\phi^0_{33}$ leaves $W_P$ unbroken. Indeed, $\phi^0_{33}$ transforms
under $U(1)_{2 \beta T_{8L}} \otimes W'_{P}$ as,
\bea
\phi^0_{33} \rightarrow \phi_{33}^{0\prime} = e^{i\frac{2}{3}(\omega' -m \pi) (1+2q)} \phi_{33}^{0 },
\label{tra1}\eea
which is invariant if $\omega' = \pi (m+\frac{3k}{1+2q})$ with
$ k=0, \pm 1, \pm 2...$. Choosing $k=0$, the residual symmetry
coincides with $W_P$ after spin parity is included and taking
$m=3$. Lastly, note that the hypercharge is
\begin{equation}
Y= \beta T_{8L}    +\frac{\sqrt{3}\beta-1}{4}(\sqrt{3}T_{8R}-T_{3R})+X',
\end{equation} 
and the electric charge is $Q=T_{3L}+Y$, all of which have the usual
form.

\subsubsection{Case 3: $w,w^\prime \sim \Lambda$}

Another possible breaking pattern takes place when assuming that the
symmetry breaking of the left-right and $SU(3)_L$ symmetry occurs at
the same scale, i.e. $w,w'\sim \La$. Therefore, we have only one new
physics scale and the gauge symmetry is directly broken down to that
of the SM as,
\bc
\begin{tabular}{c} 
    $SU(3)_C\otimes SU(3)_L\otimes SU(3)_R \otimes U(1)_X$\\
    $  \downarrow \Lambda, w, w^\prime$\\
    $SU(3)_C\otimes SU(2)_L \otimes U(1)_Y \otimes W_P$\\
    $\downarrow u, u^\prime$\\ 
    $SU(3)_C\otimes U(1)_Q\otimes W_P.$
\end{tabular}
\ec 
Here, $W_P$ is the residual discrete gauge symmetry preserved by all
VEVs and has the form obtained above.

In summary, regardless of symmetry breaking scheme adopted, they all
lead to the residual conserved $W$-parity, $W_P=(-1)^{3(B-L)+2s}$,
with $B-L=2[\beta(T_{8L}+T_{8R})+X]$.  In this way, the matter parity
is a direct consequence of the gauge group and as we shall see, it
naturally leads to the existence of stable dark matter particles.

The transformation properties of the particles of the model under
$B-L$ number and $W$-parity are collected in Table \ref{tb1}.
\begin{table}[h]
    \begin{center}
        \begin{tabular}{c|ccccccccccc}
            \hline \hline
            Particle & $\nu_{a}$ &  $e_a$ & $N_{a}$ & $u_a$ & $d_a$  & $J_{\al}$ & $J_3$ & $\ph_{11}^0$ & $\phi_{12}^+$ & $\phi_{13}^{-q}$&$\phi_{21}^-$  \\ \hline
            $B-L$ & $-1$ & $-1$ & $2q$ & $\frac{1}{3}$ & $\frac{1}{3} $& $-\frac{2(1+3q)}{3}$ & $\frac{2(2+3q)}{3}$ & $0$ & $0$ & $-(1+2q)$ & $0$   \\ \hline
            $W_P$ & 1 & 1 & $P^+$ & 1 & 1 & $P^-$ & $P^+$ & 1 & 1 & $P^-$ & 1 \\
             \hline \hline
            Particle & $\phi_{31}^q$  & $\phi_{32}^{1+q}$ & $\phi_{33}^0$ & $\phi_{22}^0$ & $\phi_{23}^{-1-q}$&$\chi_1^{(-q)}$&$\chi_{2}^{-(q+1)}$&$\chi_3^0$& $\sigma_{11}^0$& $\sigma_{12}^-$& $\sigma_{13}^q$  \\ \hline
            $B-L$ & $(1+2q)$ & $(1+2q)$ & $0$ & $0$ & $-(1+2q)$&$-(1+2q)$ &$-(1+2q)$& $0$ &$-2$ & $-2$ & $-1+2q$  \\  \hline
            $W_P$ & $P^+$ & $P^+$ & 1 & 1 & $P^-$ & $P^-$ & $P^-$ & 1 & 1 & 1 & $P^+$\\
            \hline\hline
            Particle & $\sigma_{22}^{--}$&$\sigma_{23}^{q-1}$&$\sigma_{33}^{2q}$&$A$&$Z_{L,R }$&$Z_{L,R}^\prime$&$W^\pm_{L,R}$&$X_{L,R}^q$& $X_{L,R}^{-q}$&$Y_{L,R}^{q+1}$&$Y_{L,R}^{-(q+1)}$\\ \hline
            $B-L$   &$-2$ &$-1+2q$ & $4q$ &$0$& $0$ & $0$&$0$&$1+2q$& $-(1+2q)$&$1+2q$&$-(1+2q)$\\ \hline
            $W_P$ & 1 & $P^+$ & $P^+$ & 1 & 1 & 1 & 1 & $P^+$ & $P^-$ & $P^+$ & $P^-$\\
            \hline\hline 
        \end{tabular}
    \end{center}
    \caption{\label{tb1} The $B-L$ number and $W$-parity of the model particles, with $P^\pm\equiv (-1)^{\pm(6q+1)}$.}
\end{table}

Notice that the $B-L$ charge for the new particles depends on their
electric charge, {\it i.e.} on the basic electric charge parameter
$q$, with $W$-parity values $P^\pm\equiv (-1)^{\pm(6q+1)}$. When the
new particles have ordinary electric charges $q=m/3$ for $m$ integer,
they are $W$-odd, $P^\pm=-1$, analogously to superparticles in
supersymmetry. Generally, assuming that $q\neq (2m-1)/6$, $W$-parity
is nontrivial, with $P^\pm\neq 1$ and $(P^+)^\dagger = P^-$. Such new
particles, denoted as $W$-particles in what follows, have different
$B-L$ numbers than those of the standard model. Recall that $W$-parity
is only trivial for $q=(2m-1)/6=\pm1/6,\pm1/2,\pm 5/6,\pm 7/6,\cdots$,
values not studied in this work as they require fractional charges for
the new leptons.

Since the $W$-charged and SM particles are unified within the gauge
multiplets, $W$-parity separates them into two classes,
\begin{itemize}  
\item Normal particles with $W_P=1$: Consist on all SM particles plus
  extra new fields. Explicitly, the particles belonging to this class
  are the fermions, $\nu_a$, $e_a$, $u_a$, $d_a$, the scalars,
  $\phi_{11}^0, \phi_{12}^\pm, \phi_{21}^\pm, \phi_{22}^0,
  \phi_{33}^0, $
  $ \chi_3^0, \sigma_{11}^0, \sigma_{12}^\pm, \sigma_{22}^{\pm\pm},
  \sigma_{33}^{\pm 2q}$,
  the gauge bosons, $A$, $Z_{L,R}$, $Z^\prime_{L,R}$, and the gluon.
\item $W$-particles with $W_P=P^+$ or $P^-$: Includes the new leptons
  and quarks, $N_a, J_a$, the new scalars,
  $\phi_{13}^{\pm q}, \phi_{23}^{\pm(1+q)}, \phi_{31}^{\pm q},
  \phi^{\pm (1+q)}_{32}, \chi_1^{\pm q}$,
  $\chi_2^{\pm(q+1)} \sigma^{\pm q}_{13}, \sigma_{23}^{\pm(q-1)}$, and
  the new non-Hermitian gauge bosons,
  $X_{L,R}^{\pm q}, Y_{L,R}^{\pm (q+1)}.$
\end{itemize}
It can be easily shown that $W$-particles always appear in pairs in
interactions, similarly to superparticles in supersymmetry. Indeed,
consider an interaction that includes $x$ $P^+$-fields and $y$
$P^-$-fields. The $W$-parity conservation implies
$(-1)^{(6q+1)(x-y)}=1$ for arbitrary $q$ which is satisfied only if
$x=y$. Hence, the fields $P^+$ and $P^-$ are always coupled in
pairs. The lightest $W$-particle (often called LWP) cannot decay due
to the $W$-parity conservation. Thus, if the lightest $W$-particle
carries no electrical and color charges, it can be identified as a
dark matter candidate.
   
From Table \ref{tb1}, the colorless $W$-particles have electrical
charges $\pm q, \pm(1+q), \pm(q-1)$, and therefore three dark matter
models can be built, corresponding to $q=0, \pm 1$ \footnote{The $q=1$
  case might be ruled out in the manifest left-right model
  \cite{Huong:2016kpa}.}.  The model $q=0$ includes three dark matter
candidates, namely, a lepton as the lightest mixture of $N_a^0$, a
scalar as the combination of
$\phi_{13}^0, \phi_{31}^0, \chi_1^0, \sigma_{13}^0$, and a gauge boson
from the mixing of $X^{0}_{L,R}$. The model $q=-1$ contains two dark
matter candidates: a scalar composed of
$\phi_{23}^0, \phi_{32}^0, \chi_2^0$ and a gauge boson from the
lightest mixture of $Y^{0}_{L,R}$. Lastly, the model $q=1$ has only
one dark matter candidate: the scalar field $\sigma_{23}^0$.

Before closing this section, it is important to notice that the
fundamental field $\sigma^{2q}_{33}$, carrying $W$-parity $(P^+)^2$,
leads to self-interactions among three $W$-fields, if it transforms
nontrivially under this parity. However, its presence does not alter
the results and conclusions given below. See \cite{Dong:2016sat} for a
proof.

 \section{\label{mass} Identifying physical states and masses}

 The Lagrangian of the model takes the form,
 $\mathcal{L} =\mathcal{L}_{\mathrm{gauge}} +
 \mathcal{L}_{\mathrm{Yukawa}}-V$,
 where the first term contains all kinetic terms plus gauge
 interactions. The second term includes Yukawa interactions, obtained
 by \bea \mathcal{L}_{\mathrm{Yukawa}} & =& x_{ab} \bar{\psi}^c_{aR}
 \sigma_R^\dagger \psi_{bR}+x'_{ab} \bar{\psi}^c_{aL} \sigma_L^\dagger
 \psi_{bL} + y_{ab} \bar{\psi}_{aL} \phi \psi_{b R}+z_{33}\bar{Q}_{3L}
 \phi Q_{3R} +z_{\al \beta}\bar{Q}_{\al L} \phi^* Q_{\beta R}\crn && +
 \fr{t_{3\al}}{M} \bar{Q}_{3L}\phi \chi^* Q_{\al R} + \fr{t_{\al
     3}}{M} \bar{Q}_{\al L} \phi^* \chi Q_{3R}
 +H.c., \label{yukawa}\eea where $M$ is a new physics scale that
 defines the effective interactions required to generate a consistent
 CKM matrix. The scalar potential is
 $V = V_\phi+V_\chi+V_\sigma+V_{\mathrm{mix}}$, where
 \bea V_\phi &=& \mu^2_\phi \Tr(\phi^\dagger \phi) +\la_1 [\Tr(\phi^\dagger \phi)]^2+\la_2 \Tr[(\phi^\dagger \phi)^2],\\
 V_\chi &=& \mu^2_\chi \chi^\dagger \chi + \lambda (\chi^\dagger \chi)^2,\\
 V_\sigma &=& \mu^2_\sigma \Tr(\sigma^\dagger \sigma) +\kappa_1 [\Tr(\sigma^\dagger \sigma)]^2+\kappa_2 \Tr[(\sigma^\dagger \sigma)^2],\\
 V_{\mathrm{mix}} &=& \zeta_1 \chi^\dagger \chi\Tr(\phi^\dagger
 \phi)+\zeta_2 \Tr(\phi^\dagger \phi)\Tr(\sigma^\dagger
 \sigma)+\zeta_3\Tr(\phi^\dagger \phi \sigma \sigma^\dagger)+\zeta_4
 \chi^\dagger \chi\Tr(\sigma^\dagger \sigma) \crn &&+\zeta_5
 \chi^\dagger \sigma \sigma^\dagger \chi +\zeta_6 \chi^\dagger
 \phi^\dagger \phi \chi+ (f \epsilon^{ijk} \epsilon_{\al \beta
   \gamma}\phi_i ^\al \phi_j^\beta \phi_k^\gamma+ H.c.).  \eea We see
 that $\phi$ has trilinear couplings. An $SU(2)_L$ doublet contained
 in $\phi$ can be made heavy by taking $f$ at the new physics
 scale. The remaining Higgs doublet in $\phi$ is light and lies in the
 weak scale, as shown below. If another bi-fundamental field $\rho$ is
 introduced in this minimal framework, coupling the third quark
 generation to the first two, there are no such soft-terms for
 arbitrary values of the $\beta$ parameter, since its $X$-charge is
 nonzero. Thus, both the Higgs doublets contained in $\rho$ would be
 light as their VEVs are in the weak scale. In order to avoid light
 scalars, the triplet $\chi$ is included in this work instead of
 $\rho$ in order to generate viable quark masses and mixings.
  
  \subsection{Fermion Sector}

  After spontaneous symmetry breaking, the fermions receive their
  masses via the Yukawa Lagrangian (\ref{yukawa}). For the up-type
  quarks and down-type quarks, the corresponding mass matrices are
  given by \bea
  M_u =-\frac{1}{\sqrt{2}}\left(\begin{array}{ccc} z_{11}u^\prime & z_{12}u^{\prime} & -\fr{t_{13}u^{\prime} w^{\prime}}{\sqrt{2}M}\\
                                  z_{21}u^\prime & z_{22}u^\prime & -\fr{t_{23}u^\prime w^\prime}{\sqrt{2}M} \\
                                  \fr{t_{31}uw^\prime}{\sqrt{2}M}&\fr{t_{32}uw^\prime}{\sqrt{2}M}&
                                                                                                   z_{33}u
   \end{array}\right), \hs M_d=-\fr{1}{\sqrt{2}}\left(\begin{array}{ccc}z_{11}u & z_{12}u & -\fr{t_{13}u w^\prime}{\sqrt{2} M} \\ z_{21}u & z_{22} u & -\fr{t_{23}u w^\prime }{\sqrt{2}M}\\
   \fr{t_{31}u^\prime w^\prime}{\sqrt{2}M}& \fr{t_{32}u^\prime w^\prime}{\sqrt{2}M} & z_{33}u^\prime \end{array}\right).   
  \eea 
  The ordinary quarks obtain consistent masses at the weak
    scale, $u,u'$.  The new physics or cut-off scale can be taken as
    at the largest breaking scale, $M\sim w'$. The scale M
    characterizing the non-renormalizable interaction is responsible
    for generating $V_{ub}$, $V_{cb}$, as well as quark CP violation,
    as required.
  
  The exotic quark, $J_3$, is a physical field by itself, with mass,
  $m_{J_3}=-\fr{z_{33}w}{\sqrt{2}}$, which is heavy, lying at the new
  physics regime. The two remaining exotic quarks, $J_\al $
  ($\al=1,2$), mix via a mass matrix, 
\bea
  M_{J_\al}=-\fr{1}{\sqrt{2}}\left(\begin{array}{cc} z_{11} w & z_{12}
      w \\ z_{21}w & z_{22} w \end{array}\right), \eea 
  and are both heavy, at the new physics regime too.
 
 The mass matrix elements for the charged leptons, 
 \bea
 [M_e]_{ab}=-\fr{1}{\sqrt{2}}y_{ab}u^\prime,
 \eea belong to the weak regime as usual. In contrast, the new leptons, $N_a$, have large masses dictated by the mass matrix
 \bea
  [M_N]_{ab}=-\fr{1}{\sqrt{2}}y_{ab}w.
 \eea
 
Neutrinos have both Dirac and Majorana masses. The mass matrix in the $(\nu_L\ \nu^c_R)$ basis can be written as
 \bea
 M_\nu = \left(\begin{array}{cc}M_L &M_D \\ M_D^T & M_R  \end{array}\right),
 \eea
 where $M_L, M_D, M_R$ are $3\times 3$ mass matrices, given by
 \bea  [M_D]_{ab}=-\fr{1}{\sqrt{2}}y_{ab}u, \hs  [M_L]_{ab}=-\sqrt{2}x'_{ab}v_L, \hs  [M_R]_{ab}=-\sqrt{2}x_{ab}\La, \label{neutrino} \eea
 with $\langle\sigma_{L11}^0 \rangle=v_L/\sqrt{2}$.
   As $v_L\ll u\ll \La$, the mass matrix (\ref{neutrino}) provides a
   realization of the full seesaw mechanism, producing small masses
   for the light neutrinos $\sim \nu_L$,
 \bea
 m_\nu =M_L-M_D M_R^{-1}M_D^T\sim u^2/\La-v_L,
 \eea 
and large masses for the mostly right-handed neutrinos $\sim \nu_R$, of order $M_R$.
 
\subsection{Scalar Sector}

Since $W$-parity is conserved, only the neutral fields carrying
$W_P=1$ can develop the VEVs given in (\ref{vev}). We expand the
fields around their VEVs as
\bea
&& \sigma = \left(\begin{array}{ccc} \frac{\Lambda +S_1+iA_1}{\sqrt{2}} & \frac{\sigma_{12}^-}{\sqrt{2}} & \frac{\sigma_{13}^q}{\sqrt{2}}\\
\frac{\sigma_{12}^-}{\sqrt{2}}& \sigma_{22}^{--}& \frac{\sigma_{23}^{q-1}}{\sqrt{2}} \\
\fr{\sigma_{13}^q}{\sqrt{2}}&\fr{\sigma_{23}^{q-1}}{\sqrt{2}}&\sigma_{33}^{2q}
 \end{array}\right), \crn
&& \phi= \left(\begin{array}{ccc}\frac{u+S_2+iA_2}{\sqrt{2}}& \phi_{12}^+ & \phi_{13}^{-q} \\ \phi_{21}^{-} & \frac{u^\prime+S_3+iA_3}{\sqrt{2}}& \phi_{23}^{-(q+1)}\\
\phi_{31}^q & \phi_{32}^{q+1}& \fr{w+S_4+iA_4}{\sqrt{2}} \end{array}\right), \label{expand} \\ 
&& \chi= \left( \begin{array}{ccc} \chi_1^{-q} \\ \chi_2^{-(q+1)}\\ \frac{w^\prime +S_5+iA_5}{ \sqrt{2}}  \end{array}\right).\nn	
 \eea
 The scalar potential can be written as
 $V=V_{min}+ V_{linear}+V_{mass}+V_{int}$, where $V_{min}$ is
 independent of the fields, and all interactions are grouped into
 $V_{int}$. $V_{linear}$ contains all the terms that depend linearly
 on the fields, and the gauge invariance requires,
 \bea
 && 2\mu_{\sigma}^2 +(u^{\prime 2}+w^2)\zeta_2+u^{\prime 2} (\zeta_2+\zeta_3)+ w^{\prime 2}\zeta_4+2(\kappa_1+\kappa_2)\Lambda^2=0, \nonumber \\
 && 2 \mu_\phi^2 +6\sqrt{2}f\fr{u^\prime}{u} w+2 \la_1(u^2+u^{\prime 2}+w^2)+(2 \la_2 u^2+\zeta_1 w^{\prime 2}+(\zeta_2+\zeta_3)\Lambda^2)=0, \nonumber \\
 &&2\mu_\phi^2+6\sqrt{2}f\fr{u}{u^\prime}w+2(\la_1+\la_2)u^{\prime 2}+(2\la_1(u^{\prime 2}+w^2)+\zeta_1 w^{\prime 2}+\zeta_2 \Lambda^2)=0, \nonumber \\ && 2\mu_{\phi}^2+6\sqrt{2}f \fr{uu^\prime}{w}+2\la_1(u^2+u^{\prime 2})+2(\la_1+\la_2)w^2+(\zeta_1+\zeta_6)w^{\prime 2}+\zeta_2 \Lambda^2=0,\nonumber \\
 && 2\mu_\chi^2+2\la w^{\prime 2}+\zeta_1(u^2+u^{\prime 2}+w^2) +
 \zeta_6 w^2 +\zeta_4 \Lambda^2=0.\eea 
 $V_{mass}$ consists of the terms that quadratically depend on the
 fields, and can be furhter decomposed as
 $V_{mass} = V_{mass}^A +V_{mass}^S+V_{mass}^{singly-charged}
 +V_{mass}^{doubly-charged}+V_{mass}^{q-charged}
 +V_{mass}^{(q+1)-charged}+V_{mass}^{(q-1)-charged}+V_{mass}^{2q-charged}$,
 which are listed in Appendix \ref{masstr}.

 The first mass term includes all pseudo-scalars
 $A_1,A_2,A_3,A_4,A_5$. From Appendix \ref{masstr}, we see that
 $A_1,A_5$ are massless and can be identified to the Goldstone bosons
 of the right-handed neutral gauge bosons,
 $\mathcal{Z}_R, \mathcal{Z}^\prime_R$, respectively. The remaining
 fields $A_2,A_3,A_4$ mix, but their mass matrix produces only one
 physical pseudo-scalar field with mass
 \bea
 \mathcal{A} &=& \frac{1}{\sqrt{u^2w^2+u^{\prime 2}w^2+u^2u^{\prime 2}}}\left[
 u^\prime w A_2+uwA_3+u^\prime u A_4\right], \nonumber \\ m^2_{\mathcal{A}}&=&\frac{[u^{\prime 2}w^2+u^2(u^{\prime 2}+w^2)][2\la_2(u^{\prime 2}-w^2)-\zeta_6 w^{\prime 2}]}{2u^2(w^2-u^{\prime 2})},
 \eea 
 which is heavy, at the $w,w'$ scale. The remaining fields are
 massless and orthogonal to $\mathcal{A}$
\bea
G_{\mathcal{Z}_L}&=& \sqrt{\fr{u^2(w^2+u^{\prime 2})}{w^2u^{\prime 2}+u^2(w^2+u^{\prime 2})}} \left\{-A_2+\fr{u^\prime w^2}{u(w^2+u^{\prime 2})} A_3+\fr{u^{\prime 2} w}{u(w^2+u^{\prime 2})} A_4\right\},\nonumber \\
G_{\mathcal{Z}^\prime_L}&=&\fr{u^\prime}{\sqrt{w^2+u^{\prime 2}}}A_3-\fr{w}{\sqrt{w^2+u^{\prime 2}}}A_4,
\eea
and can be identified with the Goldstone bosons of the neutral boson
$\mathcal{Z}_L$, analogous to the SM $Z$ boson, and the new neutral
gauge boson $\mathcal{Z}^\prime_L$.

The $V_{mass}^S$ term contains all the mass terms of the scalar
fields, $S_1,S_2,S_3,S_4,S_5$, as shown in Appendix \ref{masstr}.  The
five scalars mix through a $5 \times 5$ matrix. In general, it is not
easy to find the eigenstates. However, using the fact that
$u, v \ll w^\prime, w, \La $, one can diagonalize the mass matrix
perturbatively. At leading order, this matrix yields one massless
scalar field,
$H_1=\fr{1}{\sqrt{u^2+u^{\prime 2}}}\left(uS_2+u^\prime S_3\right)$,
and a massive scalar field,
$H_2=\fr{1}{\sqrt{u^2+u^{\prime 2}}}\left(u^\prime S_2- u S_3\right)$,
with
$m^2_{H_2}=-\fr{u^2+u^{\prime 2}}{2u^2}\left(\zeta_6 w^{\prime
    2}+2\la_2 w^2 \right)$.
The $H_1$ field obtains a mass at next-to-leading order,
$m_{H_1} \simeq O(u, u^\prime)$, and is identified with the standard
model Higgs boson. The remaining fields, $(S_1, S_4, S_5)$, are heavy
and mixed among themselves via a $3 \times 3$ matrix. In the limit,
$\La \gg w, w^\prime$, the corresponding physical fields have masses
given by
 \be
 H_3 = S_1, \hs m^2_{H_3}=\fr{1}{2}(\kappa_1+\kappa_2)\La^2, 
 \ee
 \bea
 H_4 &=& c_H S_4 -s_H S_5, \\ \hs m^2_{H_4}&=&\fr{1}{2}\left \{(\la_1+\la_2)w^2+\la w^{\prime 2}-\fr{(\zeta_2^2 w^2+\zeta_4^2 w^{\prime 2})}{4(\kappa_1+\kappa_2)}\right. \nonumber \\ &+& \left.  \sqrt{\left[
 (\la_1+\la_2)w^2-\la w^{\prime 2} +\fr{\zeta_4^2 w^{\prime 2}-\zeta_2^2 w^2}{4(\kappa_1+\kappa_2)}\right]^2+\fr{1}{4}w^2w^{\prime 2}\left(\fr{\zeta_2 \zeta_4}{\kappa_1+\kappa_2}-2(\zeta_1+\zeta_6)\right)^2 }\right\},\nonumber \eea
 \bea
H_5 &=& s_H S_4 +c_H S_5, \\ \hs m^2_{H_5}&=&\fr{1}{2}\left\{(\la_1+\la_2)w^2+\la w^{\prime 2}-\fr{(\zeta_2^2 w^2+\zeta_4^2 w^{\prime 2})}{4(\kappa_1+\kappa_2)}\right. \nonumber \\ &-& \left.  \sqrt{\left[
	(\la_1+\la_2)w^2-\la w^{\prime 2} +\fr{\zeta_4^2 w^{\prime 2}-\zeta_2^2 w^2}{4(\kappa_1+\kappa_2)}\right]^2+\fr{1}{4}w^2w^{\prime 2}\left(\fr{\zeta_2 \zeta_4}{\kappa_1+\kappa_2}-2(\zeta_1+\zeta_6)\right)^2 }\right\}, \nonumber
 \eea
 where the mixing angle $\theta_H$ is defined by the relation \be t_{2\theta_H}=\fr{ww^\prime \left\{-\fr{\zeta_2 \zeta_4}{\kappa_1+\kappa_2}+2(\zeta_1+\zeta_6) \right\}}{2\left\{-(\la_1+\la_2)w^2+\la w^{\prime 2}+\fr{\zeta_2^2 w^2-\zeta_4^2 w^{\prime 2}}{4(\kappa_1 +\kappa_2)} \right\}}.\ee 

 On the other hand, if one assumes that instead the hierarchy
 $w,w^\prime > \La$ holds, the masses and mixing of the heavy states,
 $(H_4,H_5)$ change accordingly to
 \bea
 t_{2\theta_H}&=& \fr{w w^\prime (\zeta_1 +\zeta_6)}{\la w^{\prime 2}-(\la_1+\la_2)w^2}, \nonumber \\
 m^2_{H_4} &=& \fr{1}{2} \left\{ (\la_1+\la_2)w^2+\la w^{\prime 2}+\sqrt{\left[ (\la_1+\la_2)w^2-\la w^{\prime 2}\right ]^2+w^2w^{\prime 2}(\zeta_1+\zeta_6)^2}\right\}, \nonumber \\  m^2_{H_5} &=& \fr{1}{2} \left\{ (\la_1+\la_2)w^2+\la w^{\prime 2}-\sqrt{\left [(\la_1+\la_2)w^2-\la w^{\prime 2}\right ]^2+w^2w^{\prime 2}(\zeta_1+\zeta_6)^2}\right\}.
 \eea 
 
 Turning now to the singly-charged Higgs fields, we have three fields
 plus their conjugates. The mass matrix extracted from (\ref{singly})
 yields four massless fields, which can be identified to the Goldstone
 bosons of the $W^\pm_{L,R}$ gauge bosons,
 \bea
 G^\pm_{W_L} &=& \frac{1}{\sqrt{u^2+u^{\prime 2}}}\left\{ u^\prime  \phi_{12}^\pm -u \phi_{21}^\pm\right\}, \nonumber \\
 G^\pm_{W_R}&=& \fr{1}{\sqrt{1+\fr{u^{\prime 2}}{u^2}+\fr{2(u^2+u^{\prime 2})^2 \La^2}{u^2(u^2-u^{\prime 2})^2}}}\left\{\fr{\sqrt{2}(u^2+u^{\prime 2})\La}{u(u^2-u^{\prime 2})} \sigma_{12}^\pm +\phi_{12}^\pm+\fr{u^\prime}{u}\phi_{21}^\pm \right\},
 \eea
 and two singly-charged massive Higgs fields with corresponding masses
 \bea
 H^{\pm} &=& \fr{\sqrt{2}u \La}{\sqrt{(u^2-u^{\prime 2})^2+2\La^2(u^2+u^{\prime 2})}}\left\{\fr{u^{\prime 2}-u^2}{\sqrt{2}u \La} \sigma_{12}^\pm +\phi_{12}^\pm +\fr{u^\prime}{u} \phi_{21}^\pm \right\}, \nonumber \\
 m^2_{H^\pm}& =& \fr{\{2\la_2(u^2-w^2)(u^{\prime 2}-w^2) +\zeta_6 w^2 w^{\prime 2}\}\{ (u^2-u^{\prime 2})^2+ 2(u^2+u^{\prime 2})\La^2\}}{4u^2(u^{\prime 2}-w^2)\La^2}.
 \eea
 
 There is only one doubly-charged Higgs field, $\sigma^{\pm\pm}_{22}$, and is physical by itself, with mass
 \bea m^2_{\sigma_{22}}= \fr{(u^2-u^{\prime 2})^2\left[2\la_2(u^2-w^2)(u^{\prime 2}-w^2)+\zeta_6w^2w^{\prime 2} \right]+2\kappa_2 u^2(w^2-u^{\prime 2})\La^4}{2u^2\La^2(u^{\prime 2}-w^2)}.
 \eea
 
 For $q$-charged scalars, $V^{q-charged}_{mass}$ contains the fields,
 $\phi_{13}^{\pm q}, \phi_{31}^{\pm q}, \sigma_{13}^{\pm q},
 \chi_1^{\pm q}$,
 as shown in Appendix \ref{masstr}. The spectrum in this sector
 includes four massless Goldstone bosons of the new gauge bosons
 $X_{L,R}^{\pm q}$,
 \bea
 G_{X_L}^{\pm q} &=& \fr{1}{\sqrt{u^4+w^4+u^2(w^{\prime 2}+2\La^2 -2 w^2)}}\left\{ (w^2-u^2)\phi_{13}^{\pm q}- \sqrt{2}u \La \sigma_{13}^{\pm q}+u w^{\prime }\chi_1^{\pm q}\right\}, \\ 
 G_{X_R}^{\pm q} &=& \fr{1}{\sqrt{(u^4+w^4+u^2(w^{\prime 2}+2\La^2 -2 w^2))\left(u^4+u^2(w^{\prime 2}+2\La^2 -2 w^2) +w^2(w^2+w^{\prime 2}+2\La^2)\right)}}\nonumber \\ && \times \left\{uw(w^{\prime 2}+2 \La^2) \phi_{13}^{\pm q}+\left( -u^4-w^4+u^2(2w^2-2\La^2-w^{\prime 2}) \right) \phi_{31}^{\pm q}+\sqrt{2}w \La(w^2-u^2)\sigma_{13}^{\pm q} \nonumber \right. \\ &&+ \left.(u^2-w^2)w^\prime w \chi_1^{\pm q}\right\}.
\eea
The remaining fields are massive. In the limit,
$\La,w,w^\prime \gg u, u^\prime$, their physical states are 
\bea
\mathcal{H}_1^{\pm q} &\simeq&
\fr{c_{\mathcal{\theta}_q}}{\sqrt{w^{\prime 2}+2 \La^2}}\left
  \{w^{\prime }\sigma_{13}^{\pm q}+\sqrt{2} \La \chi_1^{\pm q} \right
\}\crn
&& -\fr{s_{\mathcal{\theta}_q}}{\sqrt{(w^2+2\La^2)(w^2+w^{\prime 2}+2\La^2)}} \left\{(w^{\prime 2}+2\La^2)\phi_{31}^{\pm q} +  \sqrt{2}w \La \sigma_{13}^{\pm q}-ww^\prime \chi_1^{\pm q}\right \},  \\
\mathcal{H}_2^{\pm q} &\simeq&
\fr{s_{\mathcal{\theta}_q}}{\sqrt{w^{\prime 2}+2 \La^2}}\left
  \{w^{\prime }\sigma_{13}^{\pm q}+\sqrt{2} \La \chi_1^{\pm q} \right
\} \crn
&&+\fr{c_{\mathcal{\theta}_q}}{\sqrt{(w^2+2\La^2)(w^2+w^{\prime
      2}+2\La^2)}} \left\{(w^{\prime 2}+2\La^2)\phi_{31}^{\pm q} +
  \sqrt{2}w \La \sigma_{13}^{\pm q}-ww^\prime \chi_1^{\pm q}\right \},
\eea 
with masses 
\bea
&&m^2_{\mathcal{H}_1}= \fr{1}{4 \La^2} \left\{w^{\prime 2}(\zeta_5-2t_u^2 \zeta_6) \La^2+2\zeta_5 \La^4+\zeta_6 w^2(w^{\prime 2}-t_u^2w^{\prime 2}-2 \La^2) -2\la_2(t_u^2-1)w^2(w^2+2 \La^2)\right\}\nonumber \\ &&\qquad+\fr{1}{2\La(w^{\prime 2}+2 \La^2)} \sqrt{2w^2w^{\prime 2}(w^2+w^{\prime 2}+2 \La^2)\left[(t_u^2-1)(2\la_2 w^2+\zeta_6 w^{\prime 2})-2 \zeta_6 \La^2\right]^2+ \epsilon_{H^\pm q}}, \\
&&m^2_{\mathcal{H}_2}= \fr{1}{4 \La^2} \left\{w^{\prime
    2}(\zeta_5-2t_u^2 \zeta_6) \La^2+2\zeta_5 \La^4+\zeta_6
  w^2(w^{\prime 2}-t_u^2w^{\prime 2}-2 \La^2)
  -2\la_2(t_u^2-1)w^2(w^2+2 \La^2)\right\}\nonumber \\
&&\qquad-\fr{1}{2\La(w^{\prime 2}+2 \La^2)} \sqrt{2w^2w^{\prime
    2}(w^2+w^{\prime 2}+2 \La^2)\left[(t_u^2-1)(2\la_2 w^2+\zeta_6
    w^{\prime 2})-2 \zeta_6 \La^2\right]^2+ \epsilon_{H^\pm q}}, \eea
where \bea  t_{\mathcal{\theta}_q} &=& \fr{2\sqrt{2}w w^\prime \La(w^{\prime 2}+2 \La^2)\sqrt{ w^2+w^{\prime 2}+2\La^2}\left\{2\la_2(t_u^2-1)w^2+\zeta_6 \left((t_u^2-1)w^{\prime 2}-2\La^2 \right) \right\}}{\epsilon_{H^\pm q}},\nn\\
\epsilon_{H^\pm q} &=& (w^{\prime 2}+2 \La^2)\left[\La^2(w^{\prime
    2}+2 \La^2)\left(w^{\prime 2} (\zeta_5 +2t_u^2 \zeta_6)+2\zeta_5
    \La^2\right)\right.\crn &&\left.+\zeta_6 w^2(w^{\prime
    4}(1-t_u^2)+2t_u^2w^{\prime 2}\La^2-4 \La^4) \right. \nonumber \\
&&- \left. 2\la_2(t_u^2-1)w^2\left(w^2 w^{\prime 2}-2(w^2+w^{\prime
      2})\La^2 -4 \La^4 \right)\right].  \eea

$V_{mass}^{(q+1)-charged }$ contains the mixing terms of
$\phi_{23}^{\pm (q+1)}, \phi_{32}^{\pm (q+1)}, \chi_2^{\pm(q+1)}$. The
mass matrix extracted from (\ref{chargeqc1}) yields four massless
fields, identified with the Goldstone bosons of the new gauge bosons
$Y_{L,R}^{\pm(q+1)}$, and defined by
\bea
G^{\pm (q+1)}_{Y_{L}}&=& \fr{1}{\sqrt{u^{\prime 2}w^2 w^{\prime 4}+(u^{\prime 2}-w^2)^4 +(u^{\prime 6}-3u^{\prime 2}w^4+2w^6)w^{\prime 2}+w^4 w^{\prime 4}}} \times \nonumber \\ &&\left\{ \left[-(u^{\prime 2}-w^2)^2-w^2 w^{\prime 2}\right]\phi_{23}^{\pm (q+1)}+u^\prime w w^{\prime 2} \phi_{32}^{\pm (q+1)} +u^\prime w^\prime (u^{\prime 2}-w^2)\chi_2^{\pm(q+1)}\right\}, \nonumber \\ 
G^{\pm (q+1)}_{Y_{R}}&=&\fr{1}{\sqrt{(u^{\prime 2}-w^2)^2+w^2w^{\prime 2}}}\left\{(w^2-u^{\prime 2}) \phi_{32}^{\pm(q+1)}+ww^\prime \chi_2^{\pm(q+1)}\right\}. 
\eea  
The other physical fields are massive with corresponding masses,
\bea
\mathcal{ H}_{Y}^{\pm(q+1)}&=&\fr{1}{\sqrt{w^{\prime 2}(w^2+u^{\prime 2})+(u^{\prime 2}-w^2)^2}}\left\{w^\prime u^\prime \phi_{23}^{\pm(q+1)}+ww^\prime \phi_{32}^{\pm(q+1)}+(u^{\prime 2}-w^2)\chi_2^{\pm(q+1)}\right\}, \nonumber \\
m^2_{\mathcal{H}_{Y}}&=& \fr{\zeta_6}{2(u^{\prime 2}-w^2)}\left\{u^{\prime 4}+u^{\prime 2}(w^{\prime 2}-2w^2)+w^2(w^2+w^{\prime 2}) \right\}.
\eea

For $(q-1)$ and $2q-$charged scalars, $\sigma_{23}^{\pm(q-1)}$ and $\sigma_{33}^{\pm 2q}$ are already physical fields, with masses
\bea
m^2_{\sigma_{23}}&=&\fr{1}{4}\left\{\zeta_5 w^{\prime 2}-4\kappa_2 \La^2+\fr{(u^2-u^{\prime 2})(2u^2-u^{\prime 2}-w^2)\left[2\la_2(u^2-w^2)(u^{\prime 2}-w^2) +\zeta_6 w^2 w^{\prime 2}\right]}{u^2 \La^2(u^{\prime 2}-w^2)}\right\},\nonumber\\
m^2_{\sigma_{33}}&=&\fr{1}{2}\left\{\zeta_5 w^{\prime 2}-2\kappa_2 \La^2+\fr{(u^2-u^{\prime 2})(u^2-w^2)\left[ 2\la_2(u^2-w^2)(u^{\prime 2}-w^2)+\zeta_6 w^2 w^{\prime 2}\right]}{u^2(u^{\prime 2}-w^2) \La^2}   \right\}.
\eea

\subsection{Gauge-boson Sector}

Let us now study the physical gauge boson states and their masses. In
the non-Hermitian gauge boson sector, there are three kinds of
left-right gauge bosons,
$W_{L,R}^{\pm}, X_{L,R}^{\pm q}, Y_{L,R}^{\pm (q+1)}$.  The fields
$W_{L,R}^\pm$, which are defined as
$W_L^\pm =\frac{1}{\sqrt{2}}(A_{1L}\mp iA_{2L})$ and
$W_R^\pm =\frac{1}{\sqrt{2}}(A_{1R}\mp iA_{2R})$, mix through the mass
matrix,
\bea
\frac{g_L^2}{4} \left(\begin{array}{cc}
	u^2+u^{\prime 2}   & -2t_Ru u^\prime  \\
	-2t_Ru u^\prime   & t_R^2 (u^2+u^{\prime 2} +2 \Lambda^2) \\
\end{array}\right).
\eea
Diagonalizing this matrix, the eigenstates and masses are given by 
\bea
W_{1 }&=& c_\xi W_{L } -s_\xi W_{R },\hs m_{W_1}^2  \simeq \frac{g_L^2}{4}\left\{u^2+u^{\prime 2}-\fr{4t_R^2 u^2 u^{\prime 2}}{(t_R^2-1) (u^2+u^{\prime 2}) +2t_R^2 \Lambda^2} \right\} , \crn
W_{2 }&=& s_\xi W_{L } +c_\xi W_{R },\hs m_{W_2}^2  \simeq \frac{g_R^2}{4} \left \{u^2+u^{\prime 2}+2\La^2+\fr{4 t^2_R u^2 u^{\prime 2}}{(t_R^2-1) (u^2+u^{\prime 2}) +2t_R^2 \Lambda^2} \right\}, \eea
where $\La\gg u,u'$ and
$t_{2 \xi} =\fr{-4t_R u u^\prime}{2t_R^2 \La^2+(t_R^2-1)(u^2+u^{\prime
    2})}$
and $t_R =\fr{g_R}{g_L}$. $W_1$ is identified as the SM $W$ boson,
which implies $u^2+u'^2\simeq (246\ \mathrm{GeV})^2$. $W_2$ is a
physical heavy state, with mass at the new physics scale.

The mass matrix of the fields $X^{\pm q}_L =\fr{1}{\sqrt{2}}(A_{4L}\mp iA_{5L})$ and $X^{\pm q}_R =\fr{1}{\sqrt{2}}(A_{4R}\mp iA_{5R})$ is
\bea
\frac{g_L^2}{4} \left(\begin{array}{cc}
	u^2+w^2  & -2t_Ruw \\
	-2t_Ruw& t_R^2 (u^2+w^{\prime 2}+w^2+2 \Lambda^2) \\
\end{array}\right),
\eea
and yields two physical heavy states with masses 
\bea
X^{\pm q}_1 &\simeq& c_{\xi_1} X_L^\pm -s_{\xi_1} X_R^{\pm q},\\ 
m_{X_1}^2  &\simeq& \frac{g_L^2}{4} \left\{ u^2+w^2 + \frac{ 4t_R^2u^2w^2}{u^2+w^2-t_R^2(u^2+w^{\prime 2}+w^2+2 \Lambda^2) }\right\},  \\ 
X^{\pm q}_2 &\simeq& s_{\xi_1} X_L^\pm +c _{\xi_1} X_R^{\pm q},\\ 
m_{X_2}^2  &\simeq& \frac{g_R^2}{4} \left\{u^2+w^2+w^{\prime 2}+2 \Lambda^2 - \frac{ 4u^2w^2}{u^2+w^2-t_R^2(u^2+w^{\prime 2}+w^2+2 \Lambda^2) }\right\},
\eea
with $t_{2 \xi_1} = \fr{4 t_R u w}{u^2+w^2- t_R^2(u^2+w^{\prime 2}+w^2 +2 \La^2)}$.

The fields, $Y^{\pm (1+q)}_L = \fr{1}{\sqrt{2}}(A_{6L} \pm iA_{7L})$ and $Y_{R}^{\pm (1+q)}=\fr{1}{\sqrt{2}} (A_{6R}\pm i A_{7R})$, have the following mass matrix
\bea 
\frac{g_L^2}{4} \left(\begin{array}{cc}
	u^{\prime 2}+w^2  & -2t_R u^\prime w \\
	-2t_R u^\prime w& t_R^2 (u^{\prime 2}+w^{\prime 2}+w^2) \\
\end{array}\right),
\label{hu1}\eea
which provides physical heavy states with masses
\bea
Y_{1}^{\pm (1+q)}&=& c_{\xi_2}Y_L^{\pm(1+q)}-s_{\xi_2}Y_R^{\pm(1+q)},\\ 
m^2_{Y_1} &\simeq& \fr{g_L^2}{4} \left \{u^{\prime 2}+w^2 +\fr{4 t_R^2 u^{\prime 2}w^2}{u^{\prime 2}+w^2 -t_R^2(u^{\prime 2}+w^{\prime 2}+w^2)}\right\}, \\
Y_{2}^{\pm (1+q)}&=& s_{\xi_2}Y_L^{\pm(1+q)}+c_{\xi_2}Y_R^{\pm(1+q)},\\ 
m^2_{Y_2} &\simeq& \fr{g_R^2}{4} \left\{ u^{\prime 2}+w^{\prime 2}+w^2 -\fr{4  u^{\prime 2}w^2}{u^{\prime 2}+w^2 -t_R^2(u^{\prime 2}+w^{\prime 2}+w^2)}\right\},
\eea
where the mixing angle $\xi_2$ satisfies
$t_{2 \xi_2} = \fr{4 t_R u^\prime w}{u^{\prime 2}+w^2 -
  t_R^2(u^{\prime 2}+w^{ \prime 2}+w^2)}$.

The neutral gauge bosons, $A_{3L},A_{3R},A_{8L}, A_{8R},B$, mix via a $5 \times 5$ mass matrix. In order to find its eigenstates, we first work with a new basis 
\bea
A &=& s_W A_{3L}+c_W \left\{\fr{t_W}{t_R}A_{3R}+\beta t_W A_{8L}+\beta \fr{t_W}{t_R}A_{8R}+\fr{t_W}{t_X}B\right\}, \crn
Z_L &=& c_W A_{3L}-s_W \left\{\fr{t_W}{t_R}A_{3R}+\beta t_W A_{8L}+\beta \fr{t_W}{t_R}A_{8R}+\fr{t_W}{t_X}B\right\}, \crn
Z^\prime_L&=& \varsigma_1  t_X t_W \beta A_{3R}-\fr{t_W}{\varsigma_1 t_X t_R}A_{8L}+\varsigma_1  t_X t_W \beta^2 A_{8R}+\varsigma_1 t_R  t_W \beta B, \crn
Z_R&=&-\fr{\varsigma_1}{\varsigma}A_{3R}+\varsigma \varsigma_1  t_X^2 \beta A_{8R}+\varsigma \varsigma_1 t_X t_R B, \crn
Z_R^\prime &=& \varsigma (t_R A_{8R} -t_X \beta B),
\label{h2}\eea
where $t_X=\fr{g_X}{g_L}$, $ \varsigma =\fr{1}{\sqrt{t_R^2+\beta^2 t_X^2}}$, $\varsigma_1=\fr{1}{\sqrt{t_R^2+(1+\beta^2)t_X^2}}$, and $ s_W = \fr{t_X t_R}{\sqrt{t_X^2(1+\beta^2)+t_R^2(1+t_X^2(1+\beta^2))}}$. 

The gauge boson $A$ is massless and decouples, therefore it is
identified with the photon field. The remaining fields,
$Z_L,Z_L^\prime, Z_R, Z_R^\prime$, mix among themselves through a
$4\times 4$ mass matrix. Given that $w,\La \gg u,u^\prime$, the mass
matrix elements that connect $Z_L$ to $ Z_L^\prime, Z_R, Z^\prime_R$
are very suppressed. The mass matrix can be diagonalized using the
seesaw formula to separate the light state $Z_L$ from the heavy ones
$Z^\prime_L, Z_R, Z_R^\prime$. Thus, the SM $Z$ boson is identified
with $Z_L$ whose mass is
$m_{Z}^2 \simeq\fr{g_L^2}{4 c_W^2}\left( u^2+u^{\prime 2}\right)$. For
the heavy neutral gauge bosons, the mass matrix elements are
proportional to the square of the $w,w^\prime, \La$ energy scales. In
the general case, it is very difficult to find the physical heavy
states. However, if there is a hierarchy between two energy scales
$w,w'$ and $\La$, we can find them. In particular, in the limit
$\La \gg w, w^\prime$, the physical heavy states are
\bea
\mathcal{Z}^\prime_L &\simeq& Z_L^\prime, \hs 
m^2_{\mathcal{Z}_L^\prime} \simeq\frac{g_L^2}{3}\frac{(1+\varsigma_1^2t_R^2t_X^2\beta^2)^2t_W^2 w^2}{\varsigma_1^2 t_R^2 t_X^2}, \\
\mathcal{Z}_R &\simeq& c_{\xi_3} Z_R -s_{\xi_3} Z_R^\prime, \hs  \mathcal{Z}_R^\prime \simeq s_{\xi_3} Z_R +c_{\xi_3} Z_R^\prime, \\  
m^2_{\mathcal{Z}_R} &\simeq&\fr{g_L^2}{3} \frac{3w^{\prime 2}[t_R^2+t_X^2(1+\beta^2)]^2+w^2\left[3t_R^4+2t_R^2t_X^2(3+\sqrt{3}\beta)+t_X^4(3+2\sqrt{3}\beta+\beta^2)\right]}{\varsigma_1^{-2}[4+(3+2\sqrt{3}\beta +\beta^2)(t_X^2/t_R^2)]}, \\
m^2_{\mathcal{Z_R^\prime}} & \simeq & \frac{g_L^2}{3}\left\{ 4t_R^2+t_X^2(3+2\sqrt{3}\beta + \beta^2)\right\} \Lambda^2,
\eea
where the $Z_R$-$Z'_R$ mixing angle is  
\bea  
t_{2\xi_3} &=& \fr{2t_R\left [\sqrt{3}t_R^2+ \beta(3+\sqrt{3}\beta)t_X^2 \right ]\sqrt{t_R^2+t_X^2(1+\beta^2)}}{2t_R^4+t_R^2t_X^2(3-2\sqrt{3}\beta +\beta^2)-\beta^2(3+2\sqrt{3}\beta +\beta^2)t_X^4}.
\eea

With the physical states properly identified, we list in Appendix
\ref{tt} the most important interactions between the gauge bosons and
fermions in the model. Now we turn to thed to discussion of the dark
matter phenomenology.

\section{\label{darkmater} Dark matter} 

Despite the multitude of evidence for the existence of dark matter in
our universe, its nature remains a mystery and it is one of the most
exciting and important open questions in basic science
\cite{Queiroz:2016sxf}. In this work, we will investigate the possible
dark matter candidates in our model and discuss the relevant
observables, namely relic density and direct detection. Indirect
detection is not very relevant in our model because we will be
discussing multi--TeV scale dark matter, a regime for which indirect
dark matter detection cannot probe the thermal annihilation cross
section \cite{Acharya:2017ttl}.

We have seen that the $W$-parity symmetry is exact and unbroken by the
VEVs. Thus, the lightest neutral $W$-particle is stable and can be
potentially responsible for the observed DM relic density. For
concreteness we will study the model with $q=0$, i.e.
$\beta =-\fr{1}{\sqrt{3}}$. The neutral $W$-particles include a
fermion $N_a^0$, a vector gauge boson $X^{0}_{1,2}$, and a scalar
$ \mathcal{H}_{1,2}^0$ \footnote{The other two dark matter models with
  $q=\pm 1$ can be examined in a similar way.}.

\subsection{Scalar Dark Matter} 

\subsubsection{Relic Density}

Suppose that $\mathcal{H}_2^0$ is the lightest $W$-particle (LWP). It
cannot decay and can only be produced in pairs. The scalar dark matter
has only $s$-wave contribution to the annihilation
cross-section. Hence, the dark matter abundance can be approximated as
\bea
\Omega_{\mathcal{H}_2} h^2 \simeq \fr{0.1 \mathrm{pb}}{\langle \sigma v_{rel}\rangle},
\eea
where $\langle \sigma v_{rel}\rangle $ is the thermally averaged
cross-section times relative velocity. As our candidates are naturally
heavy at the new physics scale, the SM Higgs portal is inaccessible.
The main contribution to the cross-section times
relative velocity is determined by the direct annihilation channel
$\mathcal{H}_2^{0*} \mathcal{H}_2^0 \to H_1H_1$ or mediated by new
scalars. In the limit $\Lambda \gg w ,w^\prime \gg u, u^\prime$, the
interaction between $\mathcal{H}_2^0$ and $H_1$ is approximated as
\bea
\mathcal{L}_{\mathcal{H}_2^0-H_1}=\fr{\left[\la_2u^2+\la_1(u^2+u^{\prime 2})\right]c_{\theta_q}^2}{u^2+u^{\prime 2}}\mathcal{H}_2^0 \mathcal{H}_2^0 H_1 H_1.
\eea
It can be shown that the new Higgs portal gives a contribution of the
same magnitude as the one above. Therefore in our estimate it is
enough to consider only the
$\mathcal{H}_2^{0*} \mathcal{H}_2^0 \to H_1H_1$ contact
interaction. The average cross-section times relative velocity is
\bea
\langle \sigma v_{rel}\rangle = \fr{1}{16 \pi m_{\mathcal{H}_2}^2}\left\{\fr{\left[\la_2u^2
	+\la_1(u^2+u^{\prime 2})\right]c_{\theta_q}^2}{u^2+u^{\prime 2}}\right\}^2\left(1-\fr{\langle v^2\rangle }{2}-\fr{m^2_{H_1}}{m^2_{\mathcal{H}_2}} \right),
\eea
where the dark matter velocity $v$ satisfies
$\langle v^2\rangle =\fr{3}{2x_F}$, with
$x_F=m_{\mathcal{H}_2}/T_F\sim 20$ at the freezeout temperature
\cite{Bertone:2010zza}. Since $m^2_{\mathcal{H}_2}\gg m^2_{H_1}$, we
approximate
\bea
\langle \sigma v_{rel}\rangle =\left\{\fr{\al}{150\ \mathrm{GeV}}\right\}^2 \left\{\fr{\left[\la_2u^2
	+\la_1(u^2+u^{\prime 2})\right]c_{\theta_q}^2}{u^2+u^{\prime 2}}\right\}^2\left\{\fr{2.656\ \mathrm{TeV}}{m_{\mathcal{H}_2}}\right\}^2.
\eea
Thus, the dark matter candidate $\mathcal{H}^0_2$ reproduces the
correct relic density, $\Omega_{\mathcal{H}_2} h^2 \simeq 0.11$
\cite{Patrignani:2016xqp}, if $\langle \sigma v_{rel}\rangle \simeq 1$
pb, or
\bea
m_{\mathcal{H}_2}\simeq \fr{\left[\la_2u^2
	+\la_1(u^2+u^{\prime 2})\right]c_{\theta_q}^2}{u^2+u^{\prime 2}}\times 2.656\ \mathrm{TeV}\sim 2.5\ \mathrm{TeV},
\eea 
for scalar couplings of $\mathcal{O}(1)$, and using the fact that
$\al^2/(150\ \mathrm{GeV})^2 \simeq 1$ pb.  Furthermore, the above
condition implies
\bea
m_{\mathcal{H}_2} < 2.66(\la_1+\la_2)\ \mathrm{TeV} < 67\ \mathrm{TeV}, 
\eea
where the upper limit comes from the perturbativity bound
$\la_1, \la_2 < 4 \pi$. Therefore, the dark matter mass may be in the
range few TeVs to 67 TeV, depending on its interaction strength with
the SM Higgs boson.

\subsubsection{Direct Detection}

The detection through low energy nuclear recoils consitute a clear
signature for dark matter particles. Since no signal has been observed
thus far, stringent limits have been derived on the dark
matter-nucleon scattering cross section
\cite{Aguilar-Arevalo:2016zop,Aguilar-Arevalo:2016ndq,Civitarese:2016uuc,Fu:2016ega,Aprile:2017lqx,Aprile:2017iyp,Aprile:2017yea,Cui:2017nnn,Agnese:2017njq}.

In the scalar dark matter scenario, this scattering takes places
through the t-channel exchange of a $\mathcal{Z}^\prime_L$ and a heavy
scalar $\mathcal{H}^1_0$. This scenario is similar to the one studied
in \cite{Cogollo:2014jia}, where it has been shown that one can obey
direct detection limits from the XENON1T experiment with 2 years of
data for the dark matter masses above $3$~TeV, while reproducing the
correct relic density.

\subsection{Fermion Dark Matter}

\subsubsection{Relic Density}

Let us now assume that the LWP is one of the neutral fermions denoted
by $N$. The model predicts that $N$ is a Dirac fermion. The covariant
derivative (i.e., gauge interactions) dictates the dark matter
phenomenology. The dark matter might annihilate into SM particles via
the well known $Z^\prime$ portal with predictive observables
\cite{Alves:2013tqa,Alves:2015pea}. The relic density is governed by
s-channel annihilations into SM fermions, whose interactions are
presented in Appendix \ref{tt}. Assuming that the mixing between the
gauge boson $\mathcal{Z}^\prime_L$ and the other gauge bosons to be
small, which can be achived by taking
$\Lambda \gg w, w^\prime \gg u, u^\prime$, one finds the relic density
to be achieved either by annihilation into fermion pairs, or into
$ \mathcal{Z}^\prime_L \mathcal{Z}^\prime_L$. In Fig.\ref{queirozfig}
we show the relic density curve in green.

\subsubsection{Direct Detection}

The dark matter-nucleon scattering is mostly driven by the t-channel
exchange of the $\mathcal{Z}^\prime_L$ gauge boson. This scattering is
very efficient since it is governed simply the couplings with up and
down quarks without much freedom. Taking into account the current and
projected sensitivities on the dark matter-nucleon scattering
cross-section, one can conclude that the dark matter mass must lie in
the few TeV scale, as already investigated in
\cite{Alves:2016fqe}. Notice that this conclusion holds for a Dirac
fermion (the possibility of having a Majorana fermion has already been
ruled out by direct detection data \cite{Alves:2016fqe}).  The
Majorana dark matter case leads to an annihilation rate which is
helicity suppressed and therefore the range of parameter space that
yields the correct relic density is smaller compared to the Dirac
fermion scenario, only $\mathcal{Z}^\prime_L$ masses up to $2.5$~TeV
can reproduce the correct relic density in the $\mathcal{Z}^\prime_L$
resonance regime.  Although, LHC results based on heavy dilepton
resonance searches with $13.3 fb^{-1}$ of integrated luminosity
exclude $\mathcal{Z}^\prime_L$ masses below few $3.8$~TeV
\cite{Alves:2016fqe}, for this reason, the Majorana dark matter case
has already been ruled out.

\begin{figure}[h]
	\centering
	\includegraphics[scale=0.8]{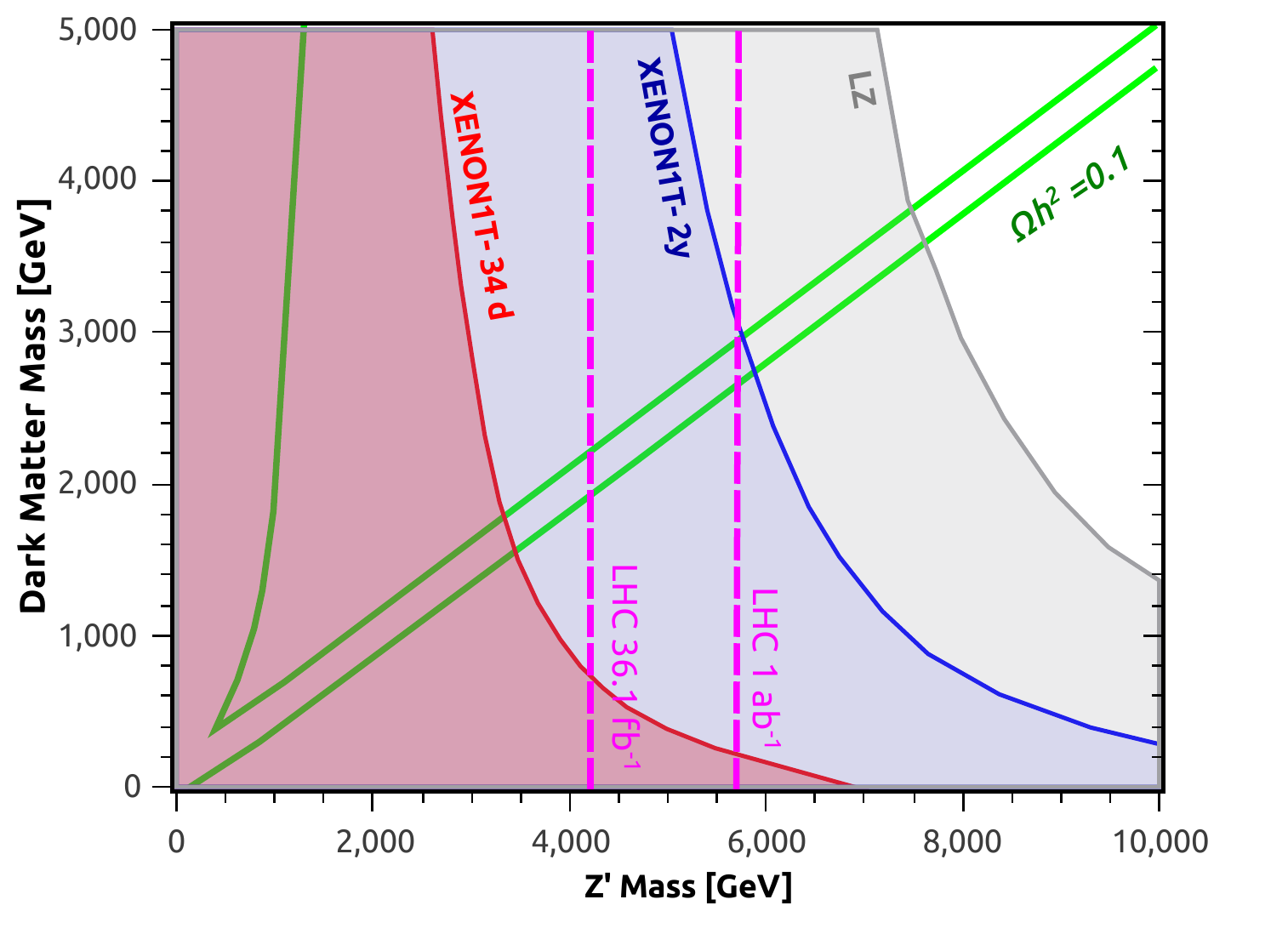}
	\caption{Summary plot for the fermion dark matter. Relic density curve (green), LHC (pink) and current direct detection limit from XENON1T-34 days (red) \cite{Aprile:2017iyp}, projected from XENON1T-2 years (blue) \cite{Aprile:2015uzo} and LZ (gray) \cite{Szydagis:2016few} are overlaid.}
	\label{queirozfig}
\end{figure}

In light of the importance of this collider bound we took the
opportunity to do a rescalling with the luminosity to obtain current
and projected limits on the $ \mathcal{Z}'_L$ mass in our model for
$36.1 fb^{-1}$ and $1000fb^{-1}$ keeping the center-of-energy of
$13$~TeV, using the collider reach tool introduced in \footnote{
  \url{http://collider-reach.web.cern.ch/?rts1=13&lumi1=3.2&rts2=13&lumi2=13.3&pdf=MSTW2008nnlo68cl.LHgrid}}. The
limits read $m_{\mathcal{Z}^\prime_L} > 4.2$~TeV and
$m_{\mathcal{Z}^\prime_L} > 5.7$~TeV, respectively. These bounds can
be seen as vertical lines in Fig.\ref{queirozfig}. We emphasize that
other limits stemming from electroweak precision or low energy physics
are subdominant thus left out of the discussion
\cite{Queiroz:2014zfa,Lindner:2016bgg}
In summary, one can conclude that our model can successfully
accommodate a Dirac fermion dark matter in agreement with existing and
projected limits near the $\mathcal{Z}^\prime_L$ resonance.

\subsection{Gauge-boson Dark Matter}

\subsubsection{Relic Density}

Finally, let us give a comment on the possibility of vector gauge
boson dark matter. In this case one assumes that the LWP is the gauge
boson $X_1^0$. It can annihilate into SM particles via following
channels,
\bea
X_1^0 X_1^{0*} \rightarrow W_1^+ W_1^-,  Z_LZ_L,  H_1 H_1, \nu \nu^c,  ll^c, qq^c,
\eea 
where $\nu=\nu_e,\nu_\mu, \nu_\tau$, $l=e, \mu, \tau$,
$q=u,d,c,s,t,b$. However, the dominant channels are
$X_1^0 X_1^{0*} \rightarrow W_1^+ W_1^-, Z_LZ_L$. Our predicted result
is similar to the one given in \cite{Dong:2013wca}. The dark matter
relic abundance is approximately given as
\bea
\Omega_{X_1} h^2 \simeq 10^{-3} \fr{m_W^2}{m_{X_1}^2}.
\eea

Since the annihilation cross section is large and it is dictated by
gauge interactions, the abundance of this vector dark matter is too
small. In the context of thermal dark matter production, the vector
dark matter in our model can contribute to only a tiny fraction of the
dark matter abundance in our universe. A similar conclusion has been
found in \cite{Mizukoshi:2010ky}.


One way to circumvent the vector dark matter underabundance is by
abdicating thermal production and tie its abundance to inflation,
where the inflaton decay or the gravitational mechanism would generate
the correct dark matter abundance \cite{Huong:2016ybt}.
  Alternatively, we mention that vector DM could be just part of
  the overall cosmological dark matter within a multicomponent thermal
  dark matter scenario.

\section{\label{conclusion} Conclusions}

We have proposed a model of flipped trinification that encompasses the
nice features of left-right and 3-3-1 models, while providing an
elegant explanation for the origin of matter parity and dark matter
stability. 
The model offers a natural framework for three types of dark matter
particles, which is an uncommon feature in UV complete models. One can
have a Dirac fermion, as well as a scalar dark matter particle, with
masses at the few TeV scale. Both scenarios reproduce the correct
relic density, while satisfying existing limits, in the context of
thermal freeze-out. As for the vector case, thermal production leads
to an under-abundant dark matter.
We have also discussed other features of the model such as the
symmetry breaking, driven by a minimal scalar content, but sufficient
to account for realistic fermion masses.
In summary, we have presented a viable theory of flipped trinification
able to account naturally for the origin of matter parity and dark
matter.

\section*{Acknowledgments}

This research is funded by Vietnam National Foundation for Science and
Technology Development (NAFOSTED) under grant numbers 103.01-2016.77
and 103.01-2017.05, and is supported by the Spanish grants
FPA2014-58183-P and SEV-2014-0398 (MINECO), and PROMETEOII/2014/084
(Generalitat Valenciana). FSQ acknowledges support from MEC and
ICTP-SAIFR FAPESP grant 2016/01343-7. C.A.V-A. acknowledges financial
support from C\'atedras CONACYT, proyect 749.

\appendix 

\section{\label{masstr} Relevant Scalar Mass Terms}

The scalar fields mix according to the class they belong, and their relevant corresponding mass terms are derived as 
\bea
V_{mass}^A&=&A_2^2\left\{\fr{w^2u^{\prime 2}(2\la_2(w^2-u^{\prime 2})+\zeta_6 w^{\prime 2})}{4u^{2}(u^{\prime 2}-w^2)} \right\}+A_2A_3 \left\{\fr{w^2u^{\prime }(2\la_2(w^2-u^{\prime 2})+\zeta_6 w^{\prime 2})}{2u(u^{\prime 2}-w^2)} \right\}\nonumber \\
&+&A_2A_4\left\{\fr{wu^{\prime 2}(2\la_2(w^2-u^{\prime 2})+\zeta_6 w^{\prime 2})}{2u(u^{\prime 2}-w^2)}\right\}+A_3^2\left\{\fr{w^2(2\la_2(w^2-u^{\prime 2})+\zeta_6 w^{\prime 2})}{4(u^{\prime 2}-w^2)} \right\} \nonumber \\
&+&A_3A_4 \left\{ \fr{wu^{\prime }(2\la_2(w^2-u^{\prime 2})+\zeta_6 w^{\prime 2})}{2(u^{\prime 2}-w^2)}\right\}+A_4^2\left\{\fr{u^{\prime 2}(2\la_2(w^2-u^{\prime 2})+\zeta_6 w^{\prime 2})}{4(u^{\prime 2}-w^2)} \right\}.
\label{pseudo}\eea
\bea
 V_{mass}^S &=&(\kappa_1+\kappa_2)\Lambda^2 S_1^2+ \left\{\frac{(u^{\prime 2}-u^2)[2 \la_2(u^2-w^2)(u^{\prime 2}-w^2)+\zeta_6 w^2 w^{\prime 2}]}{u(u^{\prime 2}-w^2)\Lambda}+\zeta_2 u \La\right\}S_1S_2+\zeta_2 u^\prime \Lambda S_1 S_3 \crn &+& \zeta_2 w \Lambda S_1S_4+\zeta_4 w^\prime \Lambda S_1S_5 +\frac{2(u^{\prime 2}-w^2)[2(\la_1+\la_2)u^4-\la_2u^{\prime 2}w^2]+\zeta_6 u^{\prime 2}w^2 w^{\prime 2}}{4u^2(u^{\prime 2}-w^2)}S_2^2 \crn &+& \fr{u^\prime\left[4\la_1 u^2+w^2 \left(2\la_2+\fr{\zeta_6 w^{\prime 2}}{w^2-u^{\prime 2}} \right)\right]}{2u}S_2S_3+\fr{w\left(4\la_1 u^2+u^{\prime 2} \left[2\la_2+\fr{\zeta_6 w^{\prime 2}}{w^2-u^{\prime 2}} \right]\right)}{2u}S_2S_4 +\zeta_1 u w^{\prime}S_2 S_5\crn &+& \left\{\la_1 u^{\prime 2}+\la_2(u^{\prime 2}-\fr{w^2}{2})-\frac{\zeta_6 w^2 w^{\prime 2}}{4(w^2-u^{\prime 2})}\right \}S_3^2+\fr{w u^\prime}{2}\left\{4 \la_1+ 2\la_2+\fr{\zeta_6 w^{\prime 2}}{w^2-u^{\prime 2}} \right \} S_3 S_4 + \zeta_1 u^\prime w^\prime S_3 S_5\crn &+&\left\{\la_1 w^2+\la_2\left(w^2-\fr{u^{\prime 2}}{2} \right) -\fr{\zeta_6 u^{\prime 2} w^{\prime 2}}{4(w^2-u^{\prime 2})}\right\}S_4^2+w w^\prime(\zeta_1+\zeta_6)S_4 S_5+ \la w^{\prime 2}S_5^2.
\label{massS} \eea
\bea
 V_{mass}^{singly-charged}&=&\fr{(u^2-u^{\prime 2})^2\left [2\la_2(u^{\prime 2}-w^2)(u^{\prime 2}-w^2)+\zeta_6w^2w^{\prime2}\right ]}{4u^2(u^{\prime 2}-w^2)\La^2} \sigma_{12}^- \sigma_{12}^+ \nonumber \crn &+& \left\{\la_2(u^{2}-w^2)+\fr{\zeta_6w^2 w^{\prime 2}}{2(u^{\prime 2}-w^2)}\right\} \phi_{12}^- \phi_{12}^++\left\{\fr{\la_2u^{\prime 2}(u^2-w^2)}{u^2} +\fr{\zeta_6 w^2 w^{\prime 2}u^{\prime 2}}{2u^2(u^{\prime 2}-w^2)}\right\}\phi_{21}^+ \phi_{21}^- \nonumber \\ &+& \left\{ \fr{(u^{\prime 2}-u^2)(2\la_2(u^{ 2}-w^2)(u^{\prime 2}-w^2)+\zeta_6w^2w^{\prime2}}{2\sqrt{2}u(u^{\prime 2}-w^2)\La} \left (\sigma_{12}^- \phi_{12}^++\fr{u^\prime}{u}\sigma_{12}^- \phi_{21}^+ \right) \right. \nonumber \\ &+& \left. \left( \fr{\la_2u^{\prime 2}(u^2-w^2)}{u^2} +\fr{\zeta_6 u^{\prime 2} w^2 w^{\prime 2}}{2 u^2(u^{\prime 2}-w^2)}\right)\phi_{12}^- \phi_{21}^++ H.c\right\}.
\label{singly} \eea
\bea
 V^{q-charged}_{mass}&=&\left\{ \la_2(u^2-u^{\prime 2})+\fr{\zeta_6u^{\prime 2}w^{\prime 2}}{2(u^{\prime 2}-w^2)}\right\}\phi_{13}^{+q} \phi_{13}^{-q} +\left\{\fr{\la_2 w^2(u^2-u^{\prime  2})}{u^2} +\fr{\zeta_6u^{\prime 2}w^{\prime 2}w^2}{2u^2(u^{\prime 2}-w^2)}\right\}\phi_{31}^{+ q} \phi_{31}^{-q} \nonumber \\ &+&\left\{\fr{(u^2-u^{\prime 2})(u^2-w^2)\left(2\la(u^2-w^2)(u^{\prime 2}-w^2)+\zeta_6 w^2 w^{\prime}+\zeta_5(u^{\prime 2}-w^2)u^2 w^{\prime 2}\La^2 \right)}{4u^2\La^2(u^{\prime 2}-w^2)} \right\}\sigma_{13}^{+q}\sigma_{13}^{-q} \nonumber\\  &+& \fr{1}{2}\left\{ \zeta_6(u^2-w^2)+\zeta_5 \La^2\right\} \chi_1^{+q}\chi_1^{-q}+ \left\{\left(\fr{\la_2 w(u^2-u^{\prime  2})}{u} +\fr{\zeta_6u^{\prime 2}w^{\prime 2}w}{2u(u^{\prime 2}-w^2)}\right)\phi_{13}^{+q} \phi_{31}^{-q} \right.  \nonumber \\ &+& \left. \fr{(u^{\prime 2}-u^2)\left (2\la_2(u^2-w^2)(u^{\prime 2}-w^2)+\zeta_6 w^2 w^{\prime 2} \right)}{2\sqrt{2}u(u^{\prime 2}-w^2) \La} \phi_{13}^{+q} \sigma_{13}^{-q}+\fr{\zeta_6 u w^\prime}{2}\phi_{13}^{+q} \chi_{1}^{-q}+\fr{\zeta_5 w^\prime \La}{2\sqrt{2}} \sigma_{13}^{+q}\chi_1^{-q} \right. \nonumber \\ &+& \left. \fr{(u^{\prime 2}-u^2)w\left (2\la_2(u^2-w^2)(u^{\prime 2}-w^2)+\zeta_6 w^2 w^{\prime 2} \right)}{2\sqrt{2}u^2(u^{\prime 2}-w^2) \La} \phi_{31}^{+q} \sigma_{13}^{-q}+\fr{\zeta_6 w w^\prime}{2}\phi_{31}^{+q}\chi_1^{-q}+ H.c.  \right\}. \label{massq}
 \eea
 \bea
V_{mass}^{(q+1)-charged}&=&\fr{\zeta_6 u^{\prime 2} w^{\prime 2}}{2(u^{\prime 2}-w^2)}\phi_{23}^{q+1}\phi_{23}^{-(q+1)} +\fr{\zeta_6 w^2 w^{\prime 2}}{2(u^{\prime 2}-w^2)}\phi_{32}^{q+1}\phi_{32}^{-(q+1)}+\fr{1}{2} \zeta_6(u^{\prime 2}-w^2)\chi_2^{(q+1)}\chi_2^{-(q+1)}\nonumber  \\ &+& \left\{ \fr{\zeta_6 u^{\prime }w w^{\prime 2}}{2(u^{\prime 2}-w^2)}\phi_{23}^{q+1}\phi_{32}^{-(q+1)}+\fr{\zeta_6 u^{\prime } w^{\prime }}{2}\phi_{23}^{q+1}\chi_{2}^{-(q+1)}+\fr{\zeta_6 w w^\prime }{2} \phi_{32}^{q+1}\chi_2^{-(q+1)}+ H.c. \right \}.\label{chargeqc1}
\eea

\section{\label{tt} Fermion Gauge-boson Interactions}

The gauge interactions of fermions arise from,
\bea
\mathcal{L} &\supset&  
 \bar{\Psi }i \gamma^\mu \partial_\mu \Psi -g_L \bar{\Psi }_L \gamma^\mu (P^{CC}_{L\mu}+P^{NC}_{L\mu})\Psi_L - g_R \bar{\Psi}_R \gamma^\mu (P^{CC}_{R\mu}+P^{NC}_{R\mu}) \Psi_R,
\label{fh2}  \eea
where $\Psi_L$ and $\Psi_R$ run on all left-handed and right-handed fermion multiplets, respectively, and  $
P^{CC}_{L,R}=\sum_{n=1,2,4,5,6,7} T_{nL,R}A_{nL,R },\
P^{NC}_{L,R }=T_{3L,R}A_{3L,R }+T_{8L,R}A_{8L,R }+\fr{g_X}{g_{L,R}} X_{\Psi_{L,R}}B.$

The interactions of the physical charged gauge bosons with fermions are 
\bea
\mathcal{L}_{CC} &=&J_{1W}^{-\mu}W_{1\mu}^+ +J_{2W}^{-\mu}W_{2\mu}^+ +J_{1X}^{-q\mu}X_{1\mu}^q +J_{2X}^{-q\mu}X_{2\mu}^q +J_{1Y}^{-(q+1)\mu}Y_{1\mu}^{q+1}+J_{2Y}^{-(q+1)\mu}Y_{2\mu}^{q+1}+H.c., \nonumber
\eea
where the charged currents take the form, 
\bea
&& J_{1W}^{-\mu}=-\fr{g_Lc_\xi }{\sqrt{2}} (\bar{\nu}_{aL}\gamma^\mu e_{aL}+\bar{u}_{aL}\gamma^\mu d_{aL})+\fr{g_R s_\xi }{\sqrt{2}}(\bar{\nu}_{aR}\gamma^\mu e_{aR}+\bar{u}_{aR}\gamma^\mu d_{aR}),\crn
&& J_{2W}^{-\mu}= -\fr{g_Ls_\xi}{\sqrt{2}} (\bar{\nu}_{aL}\gamma^\mu e_{aL}+\bar{u}_{aL}\gamma^\mu d_{aL})-\fr{g_Rc_\xi }{\sqrt{2}}(\bar{\nu}_{aR}\gamma^\mu e_{aR}+\bar{u}_{aR}\gamma^\mu d_{aR}),\crn
&& J_{1X}^{-q\mu}=-\fr{g_L c_{\xi_1} }{\sqrt{2}}(\bar{N}_{a L}\gamma^\mu \nu_{a L}-\bar{d}_{\alpha L}\gamma^\mu J_{\alpha L} +\bar{J}_{3 L}\gamma^\mu u_{3 L})+\fr{g_R s_{\xi_1}}{\sqrt{2}}(\bar{N}_{a R}\gamma^\mu \nu_{a R}-\bar{d}_{\alpha R}\gamma^\mu J_{\alpha R} +\bar{J}_{3 R}\gamma^\mu u_{3 R}),\crn && J_{2X}^{-q\mu} =-\fr{g_L s_{\xi_1} }{\sqrt{2}}(\bar{N}_{a L}\gamma^\mu \nu_{a L}-\bar{d}_{\alpha L}\gamma^\mu J_{\alpha L} +\bar{J}_{3 L}\gamma^\mu u_{3 L})-\fr{g_R c_{\xi_1}}{\sqrt{2}}(\bar{N}_{a R}\gamma^\mu \nu_{a R}-\bar{d}_{\alpha R}\gamma^\mu J_{\alpha R} +\bar{J}_{3 R}\gamma^\mu u_{3 R}), \crn
&& J_{1Y}^{-(q+1)\mu}=-\fr{g_L c_{\xi_2}}{\sqrt{2}}(\bar{N}_{a L}\gamma^\mu e_{a L} +\bar{u}_{\alpha L}\gamma^\mu J_{\alpha L} + \bar{J}_{3 L}\gamma^\mu d_{3 L})+\fr{g_R s_{\xi_2}}{\sqrt{2}}(\bar{N}_{a R}\gamma^\mu e_{a R} +\bar{u}_{\alpha R}\gamma^\mu J_{\alpha R} + \bar{J}_{3 R}\gamma^\mu d_{3 R}),\crn 
&& J_{2Y}^{-(q+1)\mu}=-\fr{g_L s_{\xi_2}}{\sqrt{2}}(\bar{N}_{a L}\gamma^\mu e_{a L} +\bar{u}_{\alpha L}\gamma^\mu J_{\alpha L} + \bar{J}_{3 L}\gamma^\mu d_{3 L})-\fr{g_R c_{\xi_2}}{\sqrt{2}}(\bar{N}_{a R}\gamma^\mu e_{a R} +\bar{u}_{\alpha R}\gamma^\mu J_{\alpha R} + \bar{J}_{3 R}\gamma^\mu d_{3 R}). \nonumber
\eea

The interactions of the physical neutral gauge bosons with fermions are obtained by
\bea
\mathcal{L}_{NC}&=&-g_L \bar{\Psi }_L \gamma^\mu P^{NC}_{L\mu}\Psi_L - g_R\bar{\Psi}_R \gamma^\mu P^{NC}_{R\mu}\Psi_R\crn
&=&-eQ(f)\bar{f}\gamma^\mu f A_\mu-\fr{g_L}{2c_W}\bar{f}\gamma^\mu [g_V^{Z_L}(f)-g_A^{Z_L}(f)\gamma_5]f Z_{L\mu}-\fr{g_L}{2c_W}\bar{f}\gamma^\mu [g_V^{\mathcal{Z^\prime}_L}(f)-g_A^{\mathcal{Z^\prime}_L}(f)\gamma_5]f \mathcal{Z^\prime}_{L\mu}\crn
&&-\fr{g_L}{2c_W}\bar{f}\gamma^\mu [g_V^{\mathcal{Z}_{R}}(f)-g_A^{\mathcal{Z}_{R}}(f)\gamma_5]f \mathcal{Z}_{R\mu}  -\fr{g_L}{2c_W}\bar{f}\gamma^\mu [g_V^{\mathcal{Z}_{R}^\prime}(f)-g_A^{\mathcal{Z}_{R}^\prime}(f)\gamma_5]f \mathcal{Z}_{R\mu}^\prime,
\label{hu}\eea where $f$ stands for every all the fermion fields, and $e=g_L s_W$. The vector and axial-vector couplings $g^{Z_L,\mathcal{Z}_L^\prime, \mathcal{Z}_R,\mathcal{Z}'_R}_{V,A}(f)$ are collected
in Tables \ref{ttZ}, \ref{ttZZ}, \ref{ttZ1}, and \ref{ttZ1p}. Note that at high energy $g_L=g_R$, i.e. $t_X=t_R$, due to the left-right symmetry. However, at the low energy, such relation does not hold anymore. Therefore, the couplings we provide are general, depending on both $t_X$ and $t_R$.  
\begin{table}[h]
	\bc
	\begin{tabular}{|c|c|c|}
		\hline
		$f$ & $g^{Z_L}_V(f)$ & $g^{Z_L}_A(f)$ \\ 
		\hline
		$\nu_a$ &  $\frac{\left\{t_R^2[3+(1+t_R)t_X(3+\sqrt{3}\beta)]+t_X^2[3+3\beta^2+t_R^2(3\beta^2-3-2\sqrt{3}\beta)] \right \}c_W^2}{6[t_R^2+t_X^2+(1+t_R^2)t_X^2\beta^2]}$ & $\frac{\left\{t_R^2 [3-(t_R-1)t_X(3+\sqrt{3}\beta)]+3(1+t_R^2)t_X^2(1+\beta^2)\right\}c_W^2}{6[t_R^2+t_X^2+(1+t_R^2)t_X^2\beta^2]}$   \\
		\hline
		$e_a$ & $-\fr{\left\{t_R^2[3-(1+t_R)t_X(3+\sqrt{3}\beta)]+t_X^2[3+3\beta^2+t_R^2(-3+2\sqrt{3}\beta +3\beta^2)]\right\}c_W^2}{6[t_R^2+t_X^2+(1+t_R^2)t_X^2 \beta^2]}$ &  $-\fr{\left\{t_R^2[3+(t_R-1)t_X(3+\sqrt{3}\beta]+3(t+t_R^2)t_X^2(1+\beta^2)\right\}c_W^2}{6[t_R^2+t_X^2+(1+t_R^2)t_X^2 \beta^2]}$  \\
		\hline $N_a$ & $\fr{t_Xt_R^2\left\{4\sqrt{3}\beta t_X+(1+t_R)(3+\sqrt{3}\beta) \right\}c_W^2}{6[t_R^2+t_X^2+(1+t_R^2)t_X^2\beta^2]}$ &  $-\fr{(t_R-1)t_R^2t_X(3+\sqrt{3}\beta)c_w^2}{6[t_R^2+t_X^2+(1+t_R^2)t_X^2\beta^2]}$ \\
		\hline
		$u_\al$ & $-\fr{\left\{3t_R^2+(1+t_R)t_Xt_R^2(1+\sqrt{3}\beta)+t_X^2[3+3\beta^2+t_R^2(3\beta^2+2\sqrt{3}\beta -3)]\right\}c_W^2}{6[t_R^2+t_X^2+(1+t_R^2)t_X^2 \beta^2]} $ & $-\fr{\left\{3t_R^2-(t_R-1)t_R^2t_X(1+\sqrt{3}\beta)+3(1+t_R^2)t_X^2(1+\beta^2)\right\}c_w^2}{6[t_R^2+t_X^2+(1+t_R^2)t_X^2 \beta^2]}$  \\
		\hline
		$u_3$ & $\fr{\left\{3t_R^2+t_R^2(1+t_R)t_X(\sqrt{3}\beta-1)+t_X^2[3+3\beta^2+t_R^2(3\beta^2-3-2\sqrt{3}\beta)]\right\}c_W} {6[t_R^2+t_X^2+(1+t_R^2)t_X^2 \beta^2]}$ & $\fr{\left\{3t_R^2-(t_R-1)t_R^2t_X(\sqrt{3} \beta-1)+3(1+t_R^2)t_X^2(1+\beta^2)\right\}c_W^2}{6[t_R^2+t_X^2+(1+t_R^2)t_X^2 \beta^2]}$ \\
		\hline $d_\al$ & $\fr{\left\{3t_R^2-t_R^2(1+t_R)t_X(1+\sqrt{3}\beta)+t_X^2[3+3\beta^2+t_R^2(3\beta^2-2\sqrt{3}\beta -3)]\right\}c_W^2}{6[t_R^2+t_X^2+(1+t_R^2)t_X^2 \beta^2]}$ & $\fr{\left\{3t_R^2+(t_R-1)t_R^2t_X(1+\sqrt{3}\beta)+3(1+t_R^2)t_X^2(1+\beta^2)\right\}c_W^2}{6[t_R^2+t_X^2+(1+t_R^2)t_X^2 \beta^2]}$\\ \hline $d_3$ & $\fr{\left\{-3t_R^2+t_R^2(1+t_R)t_X(\sqrt{3} \beta -1)-t_X^2[3+3\beta^2+t_R^2(3\beta^2+2\sqrt{3}\beta -3)]\right\}c_W^2}{6[t_R^2+t_X^2+(1+t_R^2)\beta^2 t_X^2]}$ & $\fr{\left\{-3t_R^2+(1-t_R)t_R^2t_X(\sqrt{3}\beta -1)-3t_X^2(1+t_R^2)(1+\beta^2)\right\}c_W^2}{6[t_R^2+t_X^2+(1+t_R^2)\beta^2 t_X^2]}$
		\\ \hline $J_{\al}$ & $-\fr{t_Xt_R^2[1+t_R+\sqrt{3}t_R \beta+\sqrt{3}(1-4t_X)\beta]c_W^2}{6[t_R^2+t_X^2+(1+t_R^2)t_X^2 \beta^2]}$ & $\fr{(t_R-1)t_R^2t_X(1+\sqrt{3}\beta)c_W^2}{6[t_R^2+t_X^2+(1+t_R^2)t_X^2 \beta^2]}$ \\ \hline
		$J_3$ & $\fr{t_R^2 t_X[4\sqrt{3}t_X \beta+(1+t_R)(\sqrt{3}\beta -1)]c_W^2}{6[t_R^2+t_X^2+(1+t_R^2)t_X^2 \beta^2]}$ &$\fr{(t_R-1)t_R^2t_X(\sqrt{3} \beta -1)c_W^2}{6[t_R^2+t_X^2+(1+t_R^2)t_X^2 \beta^2]}$ \\ \hline
	\end{tabular}
	\caption{\label{ttZ}The couplings of $Z_L$ with fermions.}  
	\ec
\end{table}
\begin{table}[h]
	\bc
	\begin{tabular}{|c|c|c|}
		\hline
		$f$ & $g^{Z_L^\prime}_V(f)$ & $g^{Z_L^\prime}_A(f)$ \\ 
		\hline
		$\nu_a$ &  $\frac{\left\{-\sqrt{3}(t_R^2+t_X^2)-3\beta t_Xt_R^2(1+t_R-t_X)-\sqrt{3}\beta^2t_X(1+t_R)(t_R^2+t_X-t_Rt_X) \right \}}{\zeta_1^{-1}t_R^{-1}t_X^{-1}c_W^{-1}t_W6[t_R^2+t_X^2+(1+t_R^2)t_X^2\beta^2]}$  & $\frac{\left\{t_R^3t_X\beta(3+\sqrt{3}\beta)-\sqrt{3}t_X^2(1+\beta^2)-t_R^2[\sqrt{3}+t_X(1+t_X)\beta(3+\sqrt{3}\beta)] \right \}}{\zeta_1^{-1}t_R^{-1}t_X^{-1}c_W^{-1}t_W6[t_R^2+t_X^2+(1+t_R^2)t_X^2\beta^2]}$    \\
		\hline
		$e_a$ & $\frac{\left\{-\sqrt{3}t_R^2-t_R^2(1+t_R)t_X\beta(3+\sqrt{3}\beta)+t_X^2[t_R^2\beta(-3+\sqrt{3}\beta)-\sqrt{3}(1+\beta^2)]\right \}}{\zeta_1^{-1}t_R^{-1}t_X^{-1}c_W^{-1}t_W6[t_R^2+t_X^2+(1+t_R^2)t_X^2\beta^2]}$  &  $\frac{\left\{-\sqrt{3}(t_R^2+t_X^2)+3\beta t_Xt_R^2(-1+t_R+t_X)-\sqrt{3}\beta^2t_X[t_X+t_R^2(1-t_R+t_X)] \right \}}{\zeta_1^{-1}t_R^{-1}t_X^{-1}c_W^{-1}t_W6[t_R^2+t_X^2+(1+t_R^2)t_X^2\beta^2]}$   \\
		\hline $N_a$ & $\frac{\left\{2\sqrt{3}t_R^2-\beta t_Xt_R^2(1+t_R)(3+\sqrt{3}\beta)+2\sqrt{3}t_X^2[1-(t_R^2-1)\beta^2] \right \}}{\zeta_1^{-1}t_R^{-1}t_X^{-1}c_W^{-1}t_W6[t_R^2+t_X^2+(1+t_R^2)t_X^2\beta^2]}$  &  $\frac{\left\{2\sqrt{3}t_R^2+\beta t_X t_R^2(t_R-1)(3+\sqrt{3}\beta)+2\sqrt{3}t_X^2[1+(1+t_R^2)\beta^2] \right \}}{\zeta_1^{-1}t_R^{-1}t_X^{-1}c_W^{-1}t_W6[t_R^2+t_X^2+(1+t_R^2)t_X^2\beta^2]}$  \\
		\hline
		$u_\al$ & $\frac{\left\{-\sqrt{3}t_R^2+\beta t_X t_R^2(t_R+1)(\sqrt{3}\beta+1)+t_X^2[t_R^2\beta(-3+\sqrt{3}\beta) -\sqrt{3}(1+\beta^2)] \right \}}{\zeta_1^{-1}t_R^{-1}t_X^{-1}c_W^{-1}t_W6[t_R^2+t_X^2+(1+t_R^2)t_X^2\beta^2]}$ & $\frac{\left\{-\sqrt{3}t_R^2-\beta t_X t_R^2(t_R-1)(1+\sqrt{3}\beta)-t_X^2[\beta t_R^2(\sqrt{3} \beta -3)+\sqrt{3}(1+\beta^2)] \right \}}{\zeta_1^{-1}t_R^{-1}t_X^{-1}c_W^{-1}t_W6[t_R^2+t_X^2+(1+t_R^2)t_X^2\beta^2]}$  \\
		\hline
		$u_3$ & $\frac{\left\{-\sqrt{3}t_R^2-\beta t_X t_R^2(t_R+1)(\sqrt{3}\beta-1)+t_X^2[t_R^2\beta(3+\sqrt{3}\beta) -\sqrt{3}(1+\beta^2)] \right \}}{\zeta_1^{-1}t_R^{-1}t_X^{-1}c_W^{-1}t_W6[t_R^2+t_X^2+(1+t_R^2)t_X^2\beta^2]}$ & $\frac{\left\{-\sqrt{3}(t_R^2+t_X^2)-\beta t_X t_R^2(t_R-1+3t_X)-\sqrt{3}t_X[t_X+t_R^2(1+t_X-t_R)]\beta^2 \right \}}{\zeta_1^{-1}t_R^{-1}t_X^{-1}c_W^{-1}t_W6[t_R^2+t_X^2+(1+t_R^2)t_X^2\beta^2]}$ \\
		\hline $d_\al$ & $\frac{\left\{-\sqrt{3}t_R^2+\beta t_X t_R^2(t_R+1)(\sqrt{3}\beta+1)+t_X^2[t_R^2\beta(3+\sqrt{3}\beta) -\sqrt{3}(1+\beta^2)] \right \}}{\zeta_1^{-1}t_R^{-1}t_X^{-1}c_W^{-1}t_W6[t_R^2+t_X^2+(1+t_R^2)t_X^2\beta^2]}$ & $\frac{\left\{-\sqrt{3}t_R^2-\beta t_X t_R^2(t_R-1)(1+\sqrt{3}\beta)-t_X^2[\beta t_R^2(3+\sqrt{3}\beta)+\sqrt{3}(1+\beta^2)] \right \}}{\zeta_1^{-1}t_R^{-1}t_X^{-1}c_W^{-1}t_W6[t_R^2+t_X^2+(1+t_R^2)t_X^2\beta^2]}$\\ \hline $d_3$ & $\frac{\left\{-\sqrt{3}t_R^2-\beta t_X t_R^2(t_R+1)(\sqrt{3}\beta-1)+t_X^2[t_R^2\beta(-3+\sqrt{3}\beta) -\sqrt{3}(1+\beta^2)] \right \}}{\zeta_1^{-1}t_R^{-1}t_X^{-1}c_W^{-1}t_W6[t_R^2+t_X^2+(1+t_R^2)t_X^2\beta^2]}$ & $\frac{\left\{-\sqrt{3}(t_R^2+t_X^2)+\beta t_X t_R^2(1-t_R+3t_X)-\sqrt{3}t_X[t_X+t_R^2(1+t_X-t_R)]\beta^2 \right \}}{\zeta_1^{-1}t_R^{-1}t_X^{-1}c_W^{-1}t_W6[t_R^2+t_X^2+(1+t_R^2)t_X^2\beta^2]}$
		\\ \hline $J_\al$ & $\frac{\left\{2\sqrt{3}t_R^2+\beta t_X t_R^2(t_R+1)(\sqrt{3}\beta+1)+2\sqrt{3}t_X^2[1-(t_R^2-1)\beta^2] \right \}}{\zeta_1^{-1}t_R^{-1}t_X^{-1}c_W^{-1}t_W6[t_R^2+t_X^2+(1+t_R^2)t_X^2\beta^2]}$ & $\frac{\left\{2\sqrt{3}t_R^2-\beta t_X t_R^2(t_R-1)(\sqrt{3}\beta +1)+2\sqrt{3}t_X^2[1+\beta^2(1+t_R^2)] \right \}}{\zeta_1^{-1}t_R^{-1}t_X^{-1}c_W^{-1}t_W6[t_R^2+t_X^2+(1+t_R^2)t_X^2\beta^2]}$ \\ \hline
		$J_3$ & $\frac{\left\{2\sqrt{3}t_R^2-\beta t_X t_R^2(t_R+1)(\sqrt{3}\beta-1)+2\sqrt{3}t_X^2[1-(t_R^2-1)\beta^2] \right \}}{\zeta_1^{-1}t_R^{-1}t_X^{-1}c_W^{-1}t_W6[t_R^2+t_X^2+(1+t_R^2)t_X^2\beta^2]}$ & $\frac{\left\{2\sqrt{3}t_R^2+\beta t_X t_R^2(t_R-1)(\sqrt{3}\beta -1)+2\sqrt{3}t_X^2[1+\beta^2(1+t_R^2)] \right \}}{\zeta_1^{-1}t_R^{-1}t_X^{-1}c_W^{-1}t_W6[t_R^2+t_X^2+(1+t_R^2)t_X^2\beta^2]}$ \\ \hline
	\end{tabular}
	\caption{\label{ttZZ}The couplings of $Z_L^\prime$ with fermions.}  
	\ec
\end{table}
\begin{table}[h]
	\bc
	\begin{tabular}{|c|c|}
		\hline
		$f$ & $g^\mathcal{Z_R}_V(f)$  \\
		\hline 
		$\nu_a $ & $-\fr{\left\{c_{\epsilon_2}t_R\zeta_1^{-1}[3\zeta^{-2}-\sqrt{3}\beta t_X^2+t_X(1+t_R)(3+\sqrt{3}\beta)]+s_{\epsilon_2}\zeta_1^{-2}[\sqrt{3}t_R^2+t_X \beta(3+\sqrt{3}\beta)(1+t_R)]\right\}c_W}{6\zeta^{-1} \zeta_1^{-2}}$ \\
		\hline
		$e_a$ & $-\fr{\left\{c_{\epsilon_2}t_R\zeta_1^{-1}[-3\zeta^{-2}-\sqrt{3}\beta t_X^2+t_X(1+t_R)(3+\sqrt{3}\beta)]+s_{\epsilon_2}\zeta_1^{-2}[\sqrt{3}t_R^2+t_X \beta(3+\sqrt{3}\beta)(1+t_R)]\right\}c_W}{6\zeta^{-1} \zeta_1^{-2}}$ \\
		\hline
		$N_a$ & $-\fr{\left\{c_{\epsilon_2}t_Rt_X\zeta_1^{-1}[2\sqrt{3}t_X\beta+(t_R+1)(3+\sqrt{3}\beta)]+s_{\epsilon_2}\zeta_1^{-2}[-2\sqrt{3}t_R^2+t_X \beta(3+\sqrt{3}\beta)(t_R+1)]\right\}c_W}{6\zeta^{-1} \zeta_1^{-2}}$\\
		\hline 
		$u_\al$ & $\fr{\left\{c_{\epsilon_2}t_R \zeta_1^{-1} [3\zeta^{-2}+t_R(1+t_X)+\sqrt{3}\beta t_X(1+t_R+t_X)]+s_{\epsilon_2}\zeta_1^{-2}[-\sqrt{3}t_R^2+t_X \beta(1+\sqrt{3}\beta)(1+t_R)]\right\}c_W}{6 \zeta^{-1} \zeta_1^{-2}}$
		\\
		\hline 
			$u_3$ & $\fr{\left\{c_{\epsilon_2}t_R \zeta_1^{-1} [-3\zeta^{-2}+t_X(t_R+1)+\sqrt{3}t_X(t_X-t_R-1)\beta]-s_{\epsilon_2}\zeta_1^{-2}[\sqrt{3}t_R^2+t_X \beta(\sqrt{3}\beta-1)(t_R+1)]\right\}c_W}{6 \zeta^{-1} \zeta_1^{-2}}$  \\
		\hline 
		$d_\al$ & $\fr{\left\{c_{\epsilon_2}t_R \zeta_1^{-1} [-3\zeta^{-2}+t_X(t_R+1)+\sqrt{3}\beta t_X(t_X+t_R+1)]+s_{\epsilon_2}\zeta_1^{-2}[-\sqrt{3}t_R^2+t_X \beta(\sqrt{3}\beta+1)(t_R+1)]\right\}c_W}{6 \zeta^{-1} \zeta_1^{-2}}$\\ \hline
		$d_3$ &$\fr{\left\{c_{\epsilon_2}t_R \zeta_1^{-1} [3\zeta^{-2}+t_X(t_R+1)+\sqrt{3}\beta t_X(t_X-t_R-1)]-s_{\epsilon_2}\zeta_1^{-2}[\sqrt{3}t_R^2+t_X \beta(\sqrt{3}\beta-1)(t_R+1)]\right\}c_W}{6 \zeta^{-1} \zeta_1^{-2}}$  \\
		\hline 
		$J_\al$ & $\fr{\left\{c_{\epsilon_2}t_R t_X\zeta_1^{-1} [1+t_R+\sqrt{3}\beta(1+t_R-2t_X)]+s_{\epsilon_2}\zeta_1^{-2}[2\sqrt{3}t_R^2+t_X \beta(\sqrt{3}\beta+1)(t_R+1)]\right\}c_W}{6 \zeta^{-1} \zeta_1^{-2}}$ \\
		\hline
		$J_3$ & $\fr{\left\{c_{\epsilon_2}t_R t_X\zeta_1^{-1} [1+t_R-\sqrt{3}t_R \beta-\sqrt{3}\beta(1+2t_X)]+s_{\epsilon_2}\zeta_1^{-2}[2\sqrt{3}t_R^2+t_X \beta(1-\sqrt{3}\beta)(t_R+1)]\right\}c_W}{6 \zeta^{-1} \zeta_1^{-2}}$   \\
		\hline $f$ & $g^\mathcal{Z_R}_A(f)$  \\ \hline
		$\nu_a$  & $\fr{\left\{c_{\epsilon_2}t_R\zeta_1^{-1}[3\zeta^{-2}-\sqrt{3}\beta t_X^2+t_X(t_R-1)(3+\sqrt{3}\beta)]+s_{\epsilon_2}\zeta_1^{-2}[\sqrt{3}t_R^2+t_X \beta(3+\sqrt{3}\beta)(t_R-1)]\right\}c_W}{6\zeta^{-1} \zeta_1^{-2}}$\\ \hline
		$e_a$ & $\fr{\left\{-c_{\epsilon_2}t_R\zeta_1^{-1}[3\zeta^{-2}+\sqrt{3}\beta t_X^2+t_X(t_R-1)(3+\sqrt{3}\beta)]+s_{\epsilon_2}\zeta_1^{-2}[\sqrt{3}t_R^2+t_X \beta(3+\sqrt{3}\beta)(t_R-1)]\right\} c_W}{6\zeta^{-1} \zeta_1^{-2}}$ \\
		\hline $N_a$ & $\fr{\left\{c_{\epsilon_2}t_Rt_X\zeta_1^{-1}[2\sqrt{3}t_X\beta+(t_R-1)(3+\sqrt{3}\beta)]+s_{\epsilon_2}\zeta_1^{-2}[-2\sqrt{3}t_R^2+t_X \beta(3+\sqrt{3}\beta)(t_R-1)]\right\}c_W}{6\zeta^{-1} \zeta_1^{-2}}$ \\ \hline 	$u_\al$ & $\fr{\left\{-c_{\epsilon_2}t_R \zeta_1^{-1} [3\zeta^{-2}+t_X(t_R-1)+\sqrt{3}\beta t_X(-1+t_R+t_X)]+s_{\epsilon_2}\zeta_1^{-2}[\sqrt{3}t_R^2+t_X \beta(1+\sqrt{3}\beta)(1-t_R)]\right\}c_W}{6 \zeta^{-1} \zeta_1^{-2}}$		
		\\ \hline 	$u_3$ & $\fr{\left\{c_{\epsilon_2}t_R \zeta_1^{-1} [3\zeta^{-2}+t_X(t_R-1)(\sqrt{3}\beta-1) -\sqrt{3}\beta t_X^2]+s_{\epsilon_2}\zeta_1^{-2}[\sqrt{3}t_R^2+t_X \beta(\sqrt{3}\beta-1)(t_R-1)]\right\}c_W}{6 \zeta^{-1} \zeta_1^{-2}}$
		\\
		\hline 
		$d_\al$ & $\fr{\left\{c_{\epsilon_2}t_R \zeta_1^{-1} [3\zeta^{-2}+t_X(1-t_R)-\sqrt{3}\beta t_X(t_X+t_R-1)]+s_{\epsilon_2}\zeta_1^{-2}[\sqrt{3}t_R^2+t_X \beta(\sqrt{3}\beta+1)(1-t_R)]\right\}c_W}{6 \zeta^{-1} \zeta_1^{-2}}$ \\
		\hline 
		$d_3$ & $\fr{\left\{-c_{\epsilon_2}t_R \zeta_1^{-1} [3\zeta^{-2}+t_X(t_R-1)+\sqrt{3}\beta t_X(t_X-t_R+1)]+s_{\epsilon_2}\zeta_1^{-2}[\sqrt{3}t_R^2+t_X \beta(1-\sqrt{3}\beta)(1-t_R)]\right\}c_W}{6 \zeta^{-1} \zeta_1^{-2}}$\\
		\hline 
		$J_\al$ & $\fr{\left\{c_{\epsilon_2}t_R \zeta_1^{-1} [3\zeta^{-2}+t_X(1-t_R)-\sqrt{3}\beta(t_R+t_X-1)]+s_{\epsilon_2}\zeta_1^{-2}[\sqrt{3}t_R^2+t_X \beta(\sqrt{3}\beta+1)(1-t_R)]\right\}c_W}{6 \zeta^{-1} \zeta_1^{-2}}$ \\
		\hline $J_3$ & $\fr{\left\{c_{\epsilon_2}t_R t_X\zeta_1^{-1} [2\sqrt{3}t_X \beta+(t_R-1)(\sqrt{3}\beta-1)]+s_{\epsilon_2}\zeta_1^{-2}[-2\sqrt{3}t_R^2+t_X \beta(\sqrt{3}\beta-1)(t_R-1)]\right\}c_W}{6 \zeta^{-1} \zeta_1^{-2}}$ \\
		\hline
	\end{tabular}
	\caption{\label{ttZ1} The couplings of $\mathcal{Z}_{R}$ with fermions}
	\ec
\end{table}  
\begin{table}[h]
	\bc
	\begin{tabular}{|c|c|}
		\hline
		$f$ & $g^\mathcal{Z_R^\prime}_V(f)$  \\
		\hline 
		$\nu_a $ & $\fr{\left\{-s_{\epsilon_2}t_R\zeta_1^{-1}[3\zeta^{-2}-\sqrt{3}\beta t_X^2+t_X(1+t_R)(3+\sqrt{3}\beta)]+c_{\epsilon_2}\zeta_1^{-2}[\sqrt{3}t_R^2+t_X \beta(3+\sqrt{3}\beta)(1+t_R)]\right\}c_W}{6\zeta^{-1} \zeta_1^{-2}}$ \\
		\hline
		$e_a$ & $\fr{\left\{s_{\epsilon_2}t_R\zeta_1^{-1}[3\zeta^{-2}+\sqrt{3}\beta t_X^2-t_X(1+t_R)(3+\sqrt{3}\beta)]+c_{\epsilon_2}\zeta_1^{-2}[\sqrt{3}t_R^2+t_X \beta(3+\sqrt{3}\beta)(1+t_R)]\right\}c_W}{6\zeta^{-1} \zeta_1^{-2}}$ \\
		\hline
		$N_a$ & $-\fr{\left\{s_{\epsilon_2}t_Rt_X\zeta_1^{-1}[2\sqrt{3}t_X\beta+(t_R+1)(3+\sqrt{3}\beta)]+c_{\epsilon_2}\zeta_1^{-2}[2\sqrt{3}t_R^2-t_X \beta(3+\sqrt{3}\beta)(1+t_R)]\right\}c_W}{6\zeta^{-1} \zeta_1^{-2}}$\\
		\hline 
		$u_\al$ & $\fr{\left\{s_{\epsilon_2}t_R \zeta_1^{-1} [3\zeta^{-2}+t_X(t_R+1)+\sqrt{3}\beta t_X(1+t_R+t_X)]+c_{\epsilon_2}\zeta_1^{-2}[\sqrt{3}t_R^2-t_X \beta(1+\sqrt{3}\beta)(1-t_R)]\right\}c_W}{6 \zeta^{-1} \zeta_1^{-2}}$
		\\
		\hline 
		$u_3$ & $\fr{\left\{s_{\epsilon_2}t_R \zeta_1^{-1} [-3\zeta^{-2}+t_X(t_R+1)+\sqrt{3}t_X(t_X-t_R-1)\beta]+c_{\epsilon_2}\zeta_1^{-2}[\sqrt{3}t_R^2+t_X \beta(\sqrt{3}\beta-1)(t_R+1)]\right\}c_W}{6 \zeta^{-1} \zeta_1^{-2}}$  \\
		\hline 
		$d_\al$ & $\fr{\left\{s_{\epsilon_2}t_R \zeta_1^{-1} [-3\zeta^{-2}+t_X(t_R+1)+\sqrt{3}\beta t_X(t_X+t_R+1)]+c_{\epsilon_2}\zeta_1^{-2}[\sqrt{3}t_R^2-t_X \beta(\sqrt{3}\beta+1)(t_R+1)]\right\}c_W}{6 \zeta^{-1} \zeta_1^{-2}}$\\ \hline
		$d_3$ &$\fr{\left\{s_{\epsilon_2}t_R \zeta_1^{-1} [3\zeta^{-2}+t_X(t_R+1)+\sqrt{3}\beta t_X(t_X-t_R-1)]+c_{\epsilon_2}\zeta_1^{-2}[\sqrt{3}t_R^2+t_X \beta(\sqrt{3}\beta-1)(t_R+1)]\right\}c_W}{6 \zeta^{-1} \zeta_1^{-2}}$  \\
		\hline 
		$J_\al$ & $-\fr{\left\{-s_{\epsilon_2}t_R t_X\zeta_1^{-1} [1+t_R+\sqrt{3}\beta(1+t_R-2t_X)]+c_{\epsilon_2}\zeta_1^{-2}[2\sqrt{3}t_R^2+t_X \beta(\sqrt{3}\beta+1)(t_R+1)]\right\}c_W}{6 \zeta^{-1} \zeta_1^{-2}}$ \\
		\hline
		$J_3$ & -$\fr{\left\{-s_{\epsilon_2}t_R t_X\zeta_1^{-1} [1+t_R-\sqrt{3}t_R \beta-\sqrt{3}\beta(1+2t_X)]+c_{\epsilon_2}\zeta_1^{-2}[2\sqrt{3}t_R^2+t_X \beta(1-\sqrt{3}\beta)(t_R+1)]\right\}c_W}{6 \zeta^{-1} \zeta_1^{-2}}$   \\
		\hline $f$ & $g^\mathcal{Z_R^\prime}_A(f)$  \\ \hline
		$\nu_a$  & $\fr{\left\{s_{\epsilon_2}t_R\zeta_1^{-1}[3\zeta^{-2}-\sqrt{3}\beta t_X^2+t_X(t_R-1)(3+\sqrt{3}\beta)]-c_{\epsilon_2}\zeta_1^{-2}[\sqrt{3}t_R^2+t_X \beta(3+\sqrt{3}\beta)(t_R-1)]\right\}c_W}{6\zeta^{-1} \zeta_1^{-2}}$\\ \hline
		$e_a$ & $-\fr{\left\{s_{\epsilon_2}t_R\zeta_1^{-1}[3\zeta^{-2}+\sqrt{3}\beta t_X^2-t_X(t_R-1)(3+\sqrt{3}\beta)]+c_{\epsilon_2}\zeta_1^{-2}[\sqrt{3}t_R^2+t_X \beta(3+\sqrt{3}\beta)(t_R-1)]\right\} c_W}{6\zeta^{-1} \zeta_1^{-2}}$ \\
		\hline $N_a$ & $\fr{\left\{s_{\epsilon_2}t_Rt_X\zeta_1^{-1}[2\sqrt{3}t_X\beta+(t_R-1)(3+\sqrt{3}\beta)]+c_{\epsilon_2}\zeta_1^{-2}[2\sqrt{3}t_R^2+t_X \beta(3+\sqrt{3}\beta)(1-t_R)]\right\}c_W}{6\zeta^{-1} \zeta_1^{-2}}$ \\ \hline 	$u_\al$ &- $\fr{\left\{s_{\epsilon_2}t_R \zeta_1^{-1} [3\zeta^{-2}+t_X(t_R-1)+\sqrt{3}\beta t_X(-1+t_R+t_X)]+c_{\epsilon_2}\zeta_1^{-2}[\sqrt{3}t_R^2+t_X \beta(1+\sqrt{3}\beta)(1-t_R)]\right\}c_W}{6 \zeta^{-1} \zeta_1^{-2}}$
		
		\\ \hline 	$u_3$ & $\fr{\left\{s_{\epsilon_2}t_R \zeta_1^{-1} [3\zeta^{-2}+t_X(t_R-1)(\sqrt{3}\beta-1) -\sqrt{3}\beta t_X^2]-c_{\epsilon_2}\zeta_1^{-2}[\sqrt{3}t_R^2+t_X \beta(\sqrt{3}\beta-1)(t_R-1)]\right\}c_W}{6 \zeta^{-1} \zeta_1^{-2}}$
		\\
		\hline 
		$d_\al$ & $\fr{\left\{s_{\epsilon_2}t_R \zeta_1^{-1} [3\zeta^{-2}+t_X(1-t_R)-\sqrt{3}\beta t_X(t_X+t_R-1)]-c_{\epsilon_2}\zeta_1^{-2}[\sqrt{3}t_R^2+t_X \beta(\sqrt{3}\beta+1)(1-t_R)]\right\}c_W}{6 \zeta^{-1} \zeta_1^{-2}}$ \\
		\hline 
		$d_3$ & $-\fr{\left\{s_{\epsilon_2}t_R \zeta_1^{-1} [3\zeta^{-2}+t_X(t_R-1)+\sqrt{3}\beta t_X(t_X-t_R+1)]+c_{\epsilon_2}\zeta_1^{-2}[\sqrt{3}t_R^2+t_X \beta(1-\sqrt{3}\beta)(1+t_R)]\right\}c_W}{6 \zeta^{-1} \zeta_1^{-2}}$\\
		\hline 
		$J_\al$ & $\fr{\left\{s_{\epsilon_2}t_R t_x\zeta_1^{-1} [(1-t_R)+\sqrt{3}\beta(1-t_R+2t_X)]+c_{\epsilon_2}\zeta_1^{-2}[2\sqrt{3}t_R^2+t_X \beta(\sqrt{3}\beta+1)(t_R-1)]\right\}c_W}{6 \zeta^{-1} \zeta_1^{-2}}$ \\
		\hline $J_3$ & $\fr{\left\{s_{\epsilon_2}t_R t_X\zeta_1^{-1} [2\sqrt{3}t_X \beta+(t_R-1)(\sqrt{3}\beta-1)]-c_{\epsilon_2}\zeta_1^{-2}[-2\sqrt{3}t_R^2+t_X \beta(\sqrt{3}\beta-1)(t_R-1)]\right\}c_W}{6 \zeta^{-1} \zeta_1^{-2}}$ \\
		\hline
	\end{tabular}
	\caption{\label{ttZ1p} The couplings of $\mathcal{Z}_{R}^\prime$ with fermions}
	\ec
\end{table}  

\newpage

   

\providecommand{\href}[2]{#2}\begingroup\raggedright\endgroup

 \end{document}